\documentclass[twocolumn]{aastex62}

\usepackage{amsmath}
\usepackage{graphicx}
\usepackage{natbib}
\usepackage[scaled]{helvet}
\usepackage{epsfig}
\usepackage{url}
\usepackage{textcomp}
\usepackage{mathpazo}
\usepackage{xspace}
\usepackage{hyperref}
\usepackage{apjfonts}
\usepackage{mathrsfs}
\usepackage{verbatim}
\usepackage{color,colortbl}
\usepackage[normalem]{ulem}
\usepackage{multirow}
\usepackage{lineno}

\definecolor{Gray}{gray}{0.9}

\bibpunct{(}{)}{;}{a}{}{,}
\newcommand{\bjdtdb}{\ensuremath{\rm {BJD_{TDB}}}}
\newcommand{\feh}{\ensuremath{\left[{\rm Fe}/{\rm H}\right]}}
\newcommand{\initfeh}{\ensuremath{\left[{\rm Fe}/{\rm H}\right]_0}}

\newcommand{\teff}{\ensuremath{T_{\rm eff}}}
\newcommand{\teq}{\ensuremath{T_{\rm eq}}}

\newcommand{\fbol}{\ensuremath{F_{\rm bol}}}

\newcommand{\logg}{\ensuremath{\log g_*}}

\newcommand{\msun}{\ensuremath{\,M_\Sun}}
\newcommand{\rsun}{\ensuremath{\,R_\Sun}}
\newcommand{\lsun}{\ensuremath{\,L_\Sun}}
\newcommand{\lstar}{\ensuremath{\,L_*}}
\newcommand{\rhostar}{\ensuremath{\,\rho_*}}
\newcommand{\gstar}{\ensuremath{\,g_*}}
\newcommand{\obs}{\mathrm{obs}}
\newcommand{\tru}{\mathrm{true}}

\newcommand{\sigmasb}{\ensuremath{\,\sigma_{SB}}}

\newcommand{\mj}{\ensuremath{\,M_{\rm J}}}

\newcommand{\mplanet}{\ensuremath{\,M_{\rm P}}}
\newcommand{\rplanet}{\ensuremath{\,R_{\rm P}}}
\newcommand{\rhoplanet}{\ensuremath{\,\rho_{\rm P}}}

\newcommand{\rj}{\ensuremath{\,R_{\rm J}}}

\newcommand{\fave}{\langle F \rangle}
\newcommand{\fluxcgs}{10$^9$ erg s$^{-1}$ cm$^{-2}$}

\newcommand{\gaia}{{\it Gaia}}

\newcommand{\mstar}{\ensuremath{M_{*}}}
\newcommand{\rstar}{\ensuremath{R_{*}}}

\newcommand{\exofasttwo}{{\tt EXOFASTv2}}
\newcommand{\exofast}{{\tt EXOFAST}}

\begin{document}

\title{Beating stellar systematic error floors using transit-based densities}

\author[0000-0003-3773-5142]{Jason D.\ Eastman}
\affiliation{Center for Astrophysics \textbar \ Harvard \& Smithsonian, 60 Garden St, Cambridge, MA 02138, USA}

\author[0000-0001-8274-6639]{Hannah Diamond-Lowe}
\affiliation{National Space Institute, Technical University of Denmark, Elektrovej 328, 2800 Kgs.\ Lyngby, Denmark}

\author[0000-0002-4818-7885]{Jamie Tayar}
\affiliation{Department of Astronomy, University of Florida, Bryant Space Science Center, Stadium Road, Gainesville, FL 32611, USA}

\correspondingauthor{Jason D. Eastman}
\email{jason.eastman@cfa.harvard.edu}

\title{Beating stellar systematic error floors using transit-based densities}

\shorttitle{Reducing stellar systematics with \rhostar}
\shortauthors{Eastman}

\begin{abstract}

It has long been understood that the light curve of a transiting planet constrains the density of its host star. That fact is routinely used to improve measurements of the stellar surface gravity and has been argued to be an independent check on the stellar mass. Here we show how the stellar density can also dramatically improve the precision of the radius and effective temperature of the star. This additional constraint is especially significant when we properly account for the 4.2\% radius and 2.0\% temperature systematic errors inherited from photometric zero-points, model atmospheres, interferometric calibration, and extinction. In the typical case, we can constrain stellar radii to 3\% and temperatures to 1.75\% with our evolutionary-model-based technique. In the best real-world cases, we can infer radii to 1.6\% and temperatures to 1.1\% -- well below the systematic measurement floors -- which can improve the precision in the planetary parameters by a factor of two. We explain in detail the mechanism that makes it possible and show a demonstration of the technique for a near-ideal system, WASP-4.

We also show that both the statistical and systematic uncertainties in the parallax from \gaia \ DR3 are often a significant component of the uncertainty in \lstar \ and must be treated carefully. Taking advantage of our technique requires simultaneous models of the stellar evolution, bolometric flux (e.g., a stellar spectral energy distribution), and the planetary transit, while accounting for the systematic errors in each, as is done in EXOFASTv2. 


\end{abstract}


\section{Introduction}

Any measurement of the mass, radius, or temperature of an exoplanet depends directly on those same quantities for its host star. As a result, the exploding interest in exoplanets has rekindled a broad interest in measuring precise and accurate stellar parameters in order to derive precise and accurate planetary parameters. This has correctly given rise to the oft-repeated phrase ``know thy star, know thy planet.''

Precise parallax measurements and all-sky, precise optical photometry by \gaia \ \citep{Gaia:2016,Gaia:2018}, coupled with all-sky infrared photometry from the Two Micron All Sky Survey \citep[2MASS;][]{Skrutskie:2006} and the Wide-field Infrared Survey Explorer \citep[WISE;][]{Wright:2010} have enabled a new era of precision stellar astrophysics. Now, for nearly all exoplanet host stars, the uncertainties in stellar parameters are no longer limited by the measurements, but by the stellar evolutionary and atmospheric models, as well as their calibrations, requiring an understanding of the underlying systematic errors in those models and methods. We need great care not to be overly confident in our results, but we must also not be overly conservative, lest we fail to recognize the significance of a precise detection.

\citet{Tayar:2022} published a guide for reasonable systematic error floors by carefully enumerating the sources of systematic error, tracing fundamental calibrations back to their origins, and determining the discrepancies between different groups with different instruments and different models. They find systematic errors that are larger than uncertainties often quoted in the exoplanet community.

However, \citet{Tayar:2022} do not consider the impact of an additional constraint for transiting exoplanet hosts: the stellar density, \rhostar, measured from the transit light curve. The ability of the transit light curve to measure \rhostar was first recognized by \citet{Seager:2003} for planets in circular orbits. Planets in eccentric orbits complicate the computation and generally add additional uncertainty, but with a known eccentricity and argument of periastron (typically from Radial Velocities (RVs), but also potentially through primary and secondary transits or, in the future, astrometry), we can still derive the stellar density from a transit light curve \citep[see, e.g.,][]{Eastman:2019}.

By definition, we need only two of the parameters stellar mass (\mstar), radius (\rstar), density (\rhostar), or surface gravity (\logg) to derive the others, as these are mathematically and exactly related to one another. Then, we only need either the bolometric flux (\fbol) or stellar temperature (\teff) -- and a precise distance from \gaia\ -- to determine the other, again, by definition.

This concept is not new. \citet{Sandford:2017} measured precise stellar densities of 66 Kepler planet hosts to help characterize the star. \citet{Beatty:2017} argued, and \citet{Stevens:2018} later expanded on the idea, that, with an \rstar \ from Spectral Energy Distribution (SED) fitting and \rhostar \ from transits, we can derive an \mstar \ free from the systematic errors of stellar evolutionary models. Unfortunately, \citet{Stevens:2018} showed -- and we will confirm -- that even an optimistic statistical uncertainty in the stellar mass we can obtain this way is well above the $\sim5\%$ that is thought to be a realistic systematic uncertainty in \mstar \ from stellar models. In addition, they still must rely on systematics-dominated determinations of \fbol \ to infer the stellar radius in the first place.

Here we show an idea similar to that found in \citet{Stevens:2018}, but instead of using \rhostar \ and deriving \mstar \ with a known \rstar without an evolutionary model, we use the evolutionary model to derive a more precise \rstar \ from \rhostar. Then, with a measured \lstar, we derive a more precise \teff. Assuming the systematic error floors from \citet{Tayar:2022} on \mstar \ and \lstar \ and the transit-derived measurement of \rhostar, we simply propagate those floors to significantly more precise inferred values of \rstar \ and \teff \ than the measurement floors found by \citet{Tayar:2022} using models alone. We go through the math in \S \ref{sec:derive} and discuss systematic errors in \S \ref{sec:systematics}. We talk about best practices in \S \ref{sec:implementation}. Finally, we look at a best-case scenario in refitting WASP-4b in \S \ref{sec:wasp4}, and we discuss the implications of this work in \S \ref{sec:discussion}.

\section{Mathematical validation}
\label{sec:derive}

Throughout this paper, we explain the logic of how the constraints apply by discussing the path through the primary constraints serially. Then, when we propagate the errors, we assume that we have independent measurements of each quantity at each step. In reality, the constraints are not independent, and such a serial derivation would erroneously count each measurement multiple times. We must fit all models with all constraints simultaneously within a global model to avoid this. However, the error propagation within such a global model is complex and makes it difficult to develop an understanding of where the information comes from. The excellent agreement between the errors we derive with our simplified, analytic approach and the global, simultaneous model performed by EXOFASTv2 implies that the correlations between constraints are minimal and validates our simplification as a useful pedagogical tool to develop this critical intuition. For details about how \exofasttwo \ implements these model systematic floors within our simultaneous global model, we refer the reader to the stellar model section of \citet{Eastman:2019}.

\subsection{\mstar \ from \rstar}

First, we rederive results similar to those of \citet{Stevens:2018}, which uses \rstar \ from an SED analysis and \rhostar \ from a transit light curve to determine \mstar. We start with the definition of the stellar density,

\begin{equation}
    \label{eq:rhostar}
    \rhostar \equiv \frac{3\mstar}{4\pi\rstar^3}.
\end{equation}

\noindent Assuming Gaussian and uncorrelated errors, we can use standard error propagation techniques to show that the uncertainty in \mstar, $\sigma_{\mstar}$, is given by

\begin{equation}
    \label{eq:sigma_mstar}
    \sigma_{\mstar}^2 = \left(\frac{\partial\mstar}{\partial\rhostar}\right)^2 \sigma_{\rhostar}^2 
    + \left(\frac{\partial\mstar}{\partial\rstar}\right)^2 \sigma_{\rstar}^2
\end{equation}

\noindent where $\sigma_{\rhostar}$ is the uncertainty in the stellar density measured from transits and $\sigma_{\rstar}$ is the uncertainty in the stellar radius derived from \fbol. 

To evaluate equation \ref{eq:sigma_mstar}, we solve equation \ref{eq:rhostar} for \mstar,

\begin{equation}
    \label{eq:mstar}
    \mstar =  \frac{4\pi\rstar^3\rhostar}{3},
\end{equation}

\noindent evaluate each derivative,

\begin{equation}
    \label{eq:dmdrho}
    \frac{\partial\mstar}{\partial\rhostar} = \frac{4\pi\rstar^3}{3},
\end{equation}

\begin{equation}
    \label{eq:dmdr}
    \frac{\partial\mstar}{\partial\rstar} = 4\pi\rstar^2\rhostar,
\end{equation}

\noindent and plug into equation \ref{eq:sigma_mstar}

\begin{equation}
    \label{eq:sigma_mstar_intermediate}
    \sigma_{\mstar}^2 = \left(\frac{4\pi\rstar^3\sigma_{\rhostar}}{3}\right)^2 + \left(4\pi\rstar^2\rhostar\sigma_{\rstar}\right)^2 
\end{equation}

This equation simplifies dramatically and makes the error dependence more intuitive if we divide both sides by $\mstar^2$ \ (eq \ref{eq:mstar}) to express the uncertainties as fractions:


\begin{equation}
    \label{eq:sigma_mstar_frac}
    \left(\frac{\sigma_{\mstar}}{\mstar}\right)^2 = \left(\frac{\sigma_{\rhostar}}{\rhostar}\right)^2 + \left(\frac{3\sigma_{\rstar}}{\rstar}\right)^2 
\end{equation}

\begin{figure}
  \begin{center}
    \includegraphics[width=3.5in]{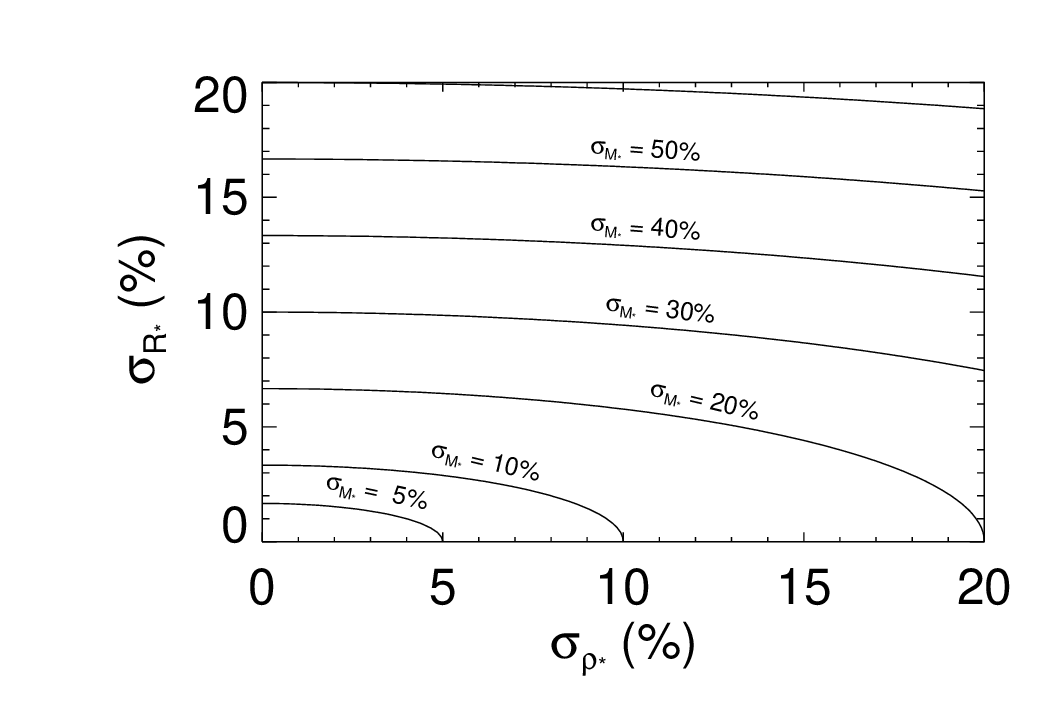} 
    \caption{A contour plot of the percent error in \mstar \ as a function of the percent errors in \rhostar \ and \rstar, derived from equation \ref{eq:sigma_mstar_frac}. We see that the dependence on $\sigma_{\rstar}$ is strong, and with a typical $\sigma_{\rhostar}$ of 10\%, the resulting constraint on \mstar \ is not particularly informative. Even with no error in \rhostar, the systematic floor in \rstar \ of $\sim4.2\%$ implies an error of 13\% in stellar mass, which is much larger than the $\sim5\%$ presumed for masses derived from evolutionary models.}
    \label{fig:mstar_errors}
  \end{center}
\end{figure}

In Figure \ref{fig:mstar_errors}, we show a contour plot of the percent error in \mstar \ as a function of the percent errors in \rhostar \ and \rstar, using equation \ref{eq:sigma_mstar_frac}. We see that the percent error in \rstar \ is multiplied by a factor of three as it propagates to the error in \mstar, which makes it difficult to get a small $\sigma_{\mstar}$ from \rstar \ and \rhostar.

\citet{Stevens:2018} do three simulated fits, one with $e=0$ (fixed), $b=0$; one with $e=0.5$, $b=0$; and one with $e=0$ (fixed), $b=0.75$ to show the dependence on the measured precision of \rhostar \ on the eccentricity and impact parameter. The percent errors from their simulations on \rstar, \mstar, \rhostar, and \teff \ are summarized in Table \ref{tab:stevens}.

\begin{deluxetable}{lcccccc}
\label{tab:stevens}
\tablecaption{Summary of relevant parameters from simulated fits by \citet{Stevens:2018}}
\tablehead{\colhead{} & \colhead{e=b=0} & \% & \colhead{e=0.5, b=0} & \% & \colhead{e=0, b=0.75} & \% }

\startdata 
\mstar   & $1.146^{+0.075}_{-0.092}$ & 7.3\% & $1.20^{+0.21}_{-0.23}$    & 18.3\% & $1.18^{+0.23}_{-0.19}$    & 17.8\% \\
\rstar   & $1.046^{+0.017}_{-0.016}$ & 1.6\% & $1.043^{+0.017}_{-0.018}$ & 1.6\%  & $1.047^{+0.018}_{-0.017}$ & 1.7\% \\
\rhostar & $1.424^{+0.049}_{-0.097}$ & 5.1\% & $1.49^{+0.23}_{-0.28}$    & 17.1\% & $1.46^{+0.26}_{-0.21}$    & 16.1\% \\
\teff    & $5710^{+160}_{-140}$      & 2.6\% & $5740^{+160}_{-150}$      & 2.7\%  & $5700^{+170}_{-140}$      & 2.7\%  \\
\enddata 
\tablecomments{Percent errors were calculated by averaging upper and lower errors.}
\end{deluxetable}

While their quoted uncertainty in \rstar \ ignores the often dominant sources of systematic error, we can still use these results to confirm our expectations given Figure \ref{fig:mstar_errors} and equation \ref{eq:sigma_mstar_frac}. Indeed, in the first case, with 5.1\% errors on \rhostar \ and 1.6\% errors on \rstar, Equation  \ref{eq:sigma_mstar} predicts a $7.0\%$ uncertainty in \mstar, completely consistent with the measured value of 7.3\%. In the second case, with 17.1\% errors on \rhostar \ and 1.6\% errors on \rstar, equation  \ref{eq:sigma_mstar}  predicts a $17.8\%$ uncertainty in \mstar, completely consistent with the measured value of 18.3\%. And finally, in the last case, with 16.1\% errors on \rhostar \ and 1.7\% errors on \rstar, we predict a $17.0\%$ uncertainty in \mstar, again consistent with the measured value of 17.8\%.

The uncertainty in \mstar, even in the best case they present and ignoring systematic errors in \fbol \ when they derive \rstar, is well above the $\sim5\%$ systematic uncertainties presumed in most stellar models because the stellar radius uncertainty is compounded, leading to a much larger percent error in the stellar mass.

For certain ideal systems (tidally circularized, deep, well measured, like WASP-4), the error on \rhostar \ can be $\sim$1\%. Even so, coupled with the 4.2\% systematics-dominated uncertainty in \rstar \ recommended by \citet{Tayar:2022}, the resultant mass uncertainty is still $\sim 13\%$. We also note that the uncertainty in \teff \ from \citet{Stevens:2018} is relatively large because they have avoided using an evolutionary model, and so the SED is working to constrain both \rstar \ and \teff.

Hence, while their method may serve as a rough, independent check on systematics, it is unlikely to be helpful in a significant number of cases.

\subsection{\rstar \ from \mstar}
\label{sec:rstar}

Instead, we can infer \rstar \ from \mstar \ and \rhostar. Again, starting from equation \ref{eq:rhostar}, we instead propagate the uncertainties in \mstar \ to \rstar.

\begin{equation}
    \label{eq:sigma_rstar}
    \sigma_{\rstar}^2 = \left(\frac{\partial\rstar}{\partial\rhostar}\right)^2 \sigma_{\rhostar}^2 
    + \left(\frac{\partial\rstar}{\partial\mstar}\right)^2 \sigma_{\mstar}^2
\end{equation}

As before, we evaluate and simplify by dividing both sides by $\rstar^2$ \ to express it as a percent uncertainty:

\begin{equation}
    \label{eq:sigma_rstar_frac}
    \left(\frac{\sigma_{\rstar}}{\rstar}\right)^2 = 
    \left(\frac{\sigma_{\rhostar}}{3\rhostar}\right)^2 + \left(\frac{\sigma_{\mstar}}{3\mstar}\right)^2
\end{equation}

Here we see that, instead of magnifying the errors as when we determined \mstar, our input fractional errors are reduced by a factor of 3. Thus, the resultant percent uncertainty in \rstar \ can be much lower than the input percent uncertainties in \mstar \ and \rhostar. 

\begin{figure}
  \begin{center}
    \includegraphics[width=3.5in]{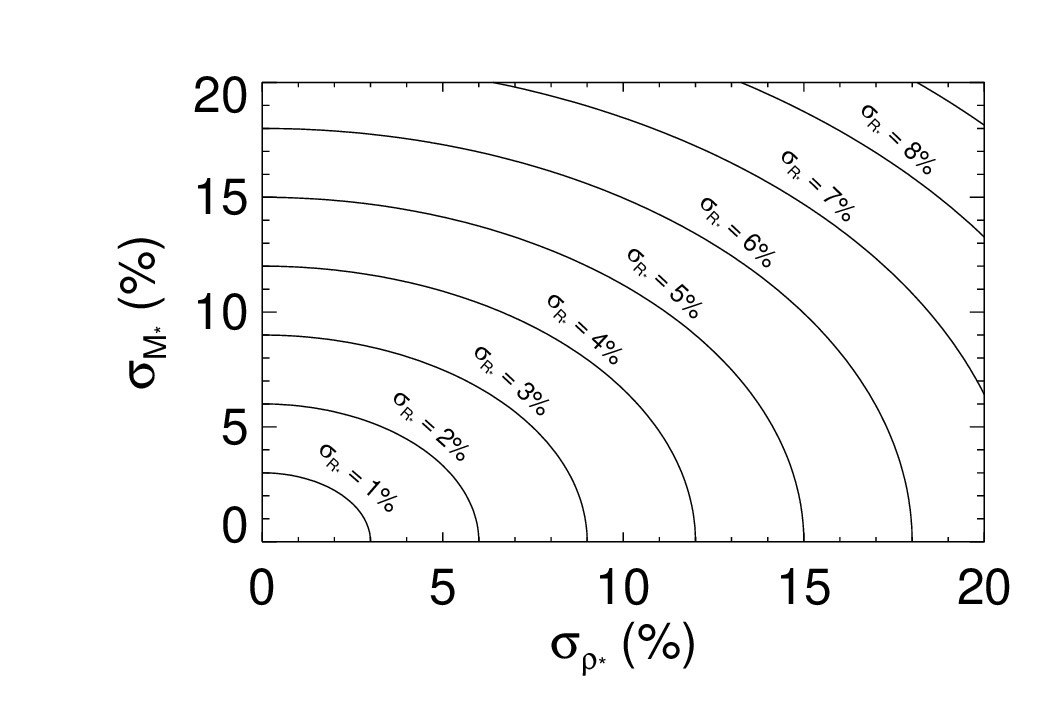}
    \caption{A contour plot of the percent error in \rstar \ as a function of the percent errors in \rhostar \ and \mstar, derived in equation \ref{eq:sigma_rstar_frac}. We see that the dependence on \mstar \ is much weaker, and with a typical $\sigma_{\rhostar}$ of 10\%, the resulting constraint on \rstar \ can be well below the systematic floor from evolutionary models. In the best cases, we can measure \rhostar \ to $\sim1\%$, resulting in \rstar \ uncertainties of $\sim1.7\%$ -- almost entirely dominated by the systematic floor in \mstar.}
    \label{fig:rstar_errors}
  \end{center}
\end{figure}

Figure \ref{fig:rstar_errors} shows the same contour plot, but now with the propagated percent error in \rstar \ as a function of percent errors in \rhostar \ and \mstar \ (equation \ref{eq:sigma_rstar_frac}). Of course, getting \mstar \ usually requires stellar evolutionary models, but even assuming a systematic floor of 5\% on \mstar \ from stellar models, if we measure \rhostar \ to 1\%, we get 1.7\% errors on \rstar -- almost three times better than the recommended systematic floor in the stellar models. And while our determination of the stellar radius hinges on the systematics-dominated stellar evolution models, these errors are already included in the computation as $\sigma_{\mstar}$. For everything else, we rely on well-established physics (i.e., Kepler's law) and definitions (e.g., equation \ref{eq:rhostar}), though it is important to understand the potential sources of systematic error in \rhostar, discussed in \S \ref{sec:systematics}.

\begin{figure}
  \begin{center}
    \includegraphics[width=3.5in]{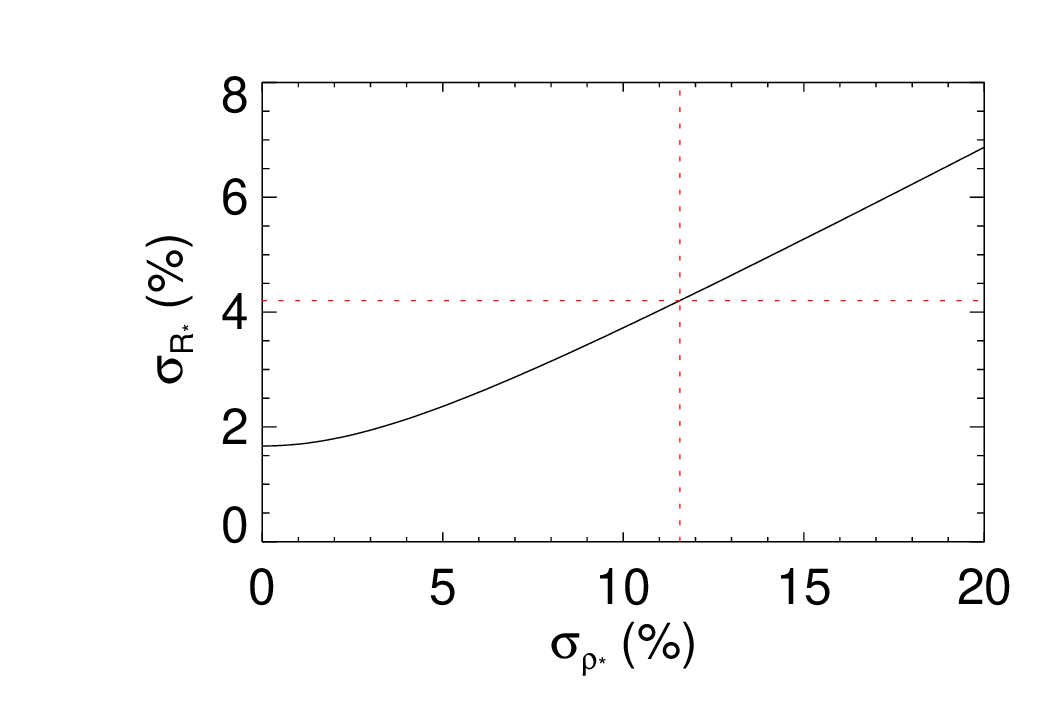}
    \caption{A plot of $\sigma_{\rstar}$ as a function of $\sigma_{\rhostar}$, assuming $\sigma_{\mstar} = 5\%$. The break-even point, where $\sigma_{\rstar}=4.2\%$ from \citet{Tayar:2022}, is shown as a vertical red dashed line and corresponds to $\sigma_{\rhostar}\sim 11.5\%$.}
    \label{fig:rstar_vs_rhostar}
  \end{center}
\end{figure}

Assuming that our \mstar \ error is 5\%, we can plot $\sigma_{\rstar}$ as a function of $\sigma_{\rhostar}$, as shown in Figure \ref{fig:rstar_vs_rhostar}, where the break-even point is shown as a vertical red dashed line at $\sim 11.5\%$. That is, measuring \rhostar \ to better than 11.5\% -- which is typical -- allows us to measure the stellar radius to better than the systematic errors identified by \citet{Tayar:2022}. 

\subsection{\teff \ from \lstar}
\label{sec:tefffromlstar}

Now we can propagate the error in \rstar \ along with a reasonable floor in \lstar \ to \teff \ with the definition of the stellar luminosity,

\begin{equation}
    \label{eq:lstar}
    \lstar \equiv  4\pi\rstar^2\sigmasb \teff^4,
\end{equation}

\noindent where $\sigmasb$ is the Stefan-Boltzmann constant. Following a similar procedure to that above, we write

\begin{equation}
    \label{eq:sigma_teff}
    \sigma_{\teff}^2 = \left(\frac{\partial\teff}{\partial\lstar}\right)^2 \sigma_{\lstar}^2 + \left(\frac{\partial\teff}{\partial\rstar}\right)^2 \sigma_{\rstar}^2.
\end{equation}

Again, we evaluate and simplify by dividing by $\teff^2$ \ to express it in terms of fractional errors:

\begin{equation}
    \label{eq:sigma_teff_frac_lstar}
    \left(\frac{\sigma_{\teff}}{\teff}\right)^2 = \left(\frac{\sigma_{\lstar}}{4\lstar}\right)^2 + \left(\frac{\sigma_{\rstar}}{2\rstar}\right)^2
\end{equation}

We see that the fractional uncertainty in \lstar \ is cut by a factor of 4 as it propagates to \teff, so the uncertainty in \rstar \ quickly dominates. Even so, the uncertainty in \rstar \ is also halved, leading to surprisingly precise determinations of \teff.

\begin{figure}
  \begin{center}
    \includegraphics[width=3.5in]{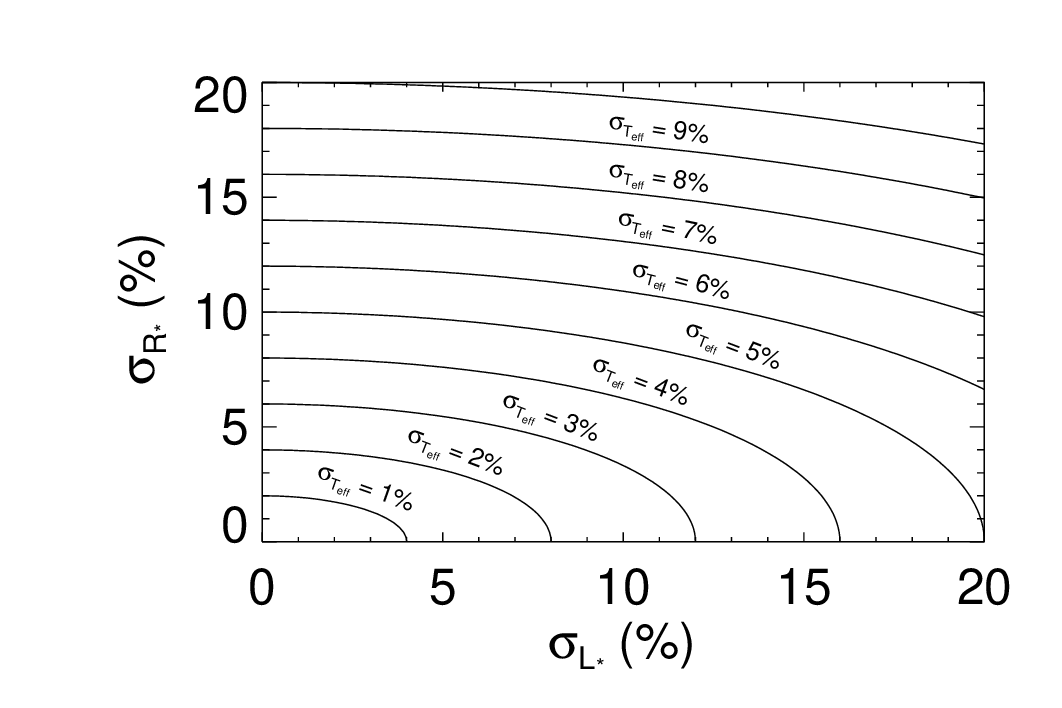} 
    \caption{A contour plot of the percent error in \teff \ as a function of the percent errors in \rstar \ and \fbol, derived from equation \ref{eq:sigma_teff}. We see that the dependence on \mstar \ is much weaker, and with a typical $\sigma_{\rstar}$ of 3\% and a systematic floor of $\lstar\sim2.4\%$, the resulting constraint on \teff \ can be $\sim 1.5\%$ -- well below the systematic floor from evolutionary models, SED models, or spectra. In the best cases, we can measure \rstar \ to $\sim1.7\%$ and \teff \ uncertainties of just under 1\%.}
    \label{fig:teff_errors}
  \end{center}
\end{figure}

Figure \ref{fig:teff_errors} shows a contour plot of equation \ref{eq:sigma_teff_frac_lstar} and we see that, for the recommended systematic error floor of 2.4\% on \lstar \ from \citet{Tayar:2022} and our best-case error on \rstar \ of 1.7\% above, we get 0.9\% errors on \teff. That is 50 K for a solar-type star -- far better than the typically assumed systematic error floors on \teff, which are derived from the complexities of calibrations and gaps in our knowledge of stellar evolution and atmospheres. Instead, these are very simply derived from an independent constraint on \rhostar \ (based on Kepler's law) \ and error propagation, with well-motivated systematic error floors on \lstar \ and \mstar \ from \citet{Tayar:2022}.

\begin{figure}
  \begin{center}
    \includegraphics[width=3.5in]{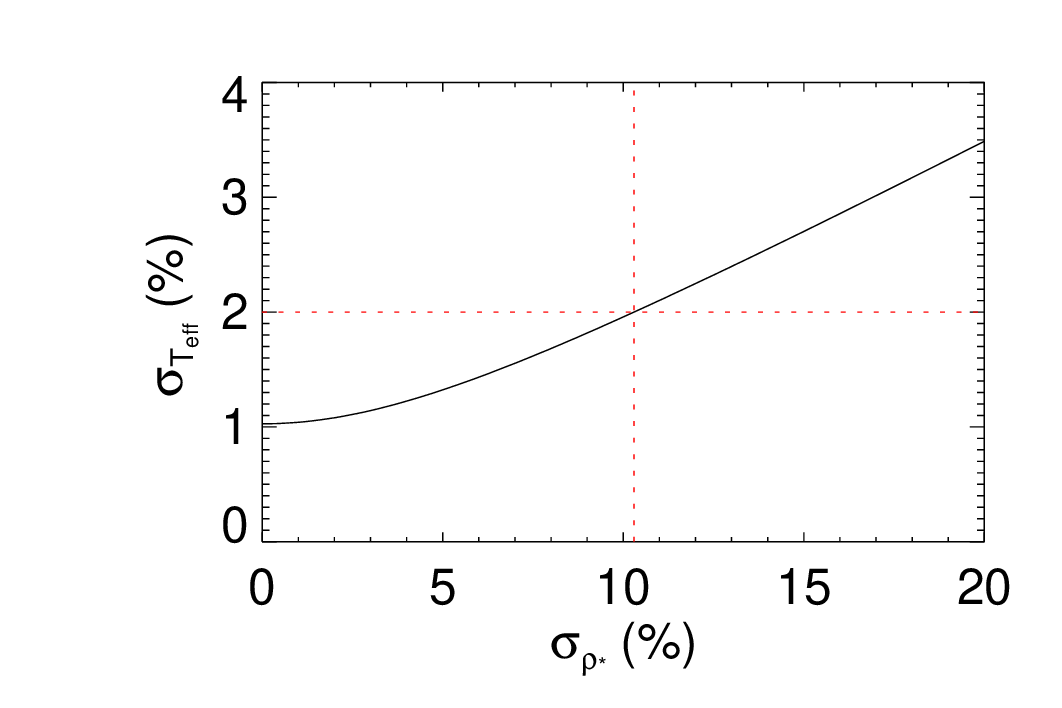}
    \caption{A plot of $\sigma_{\teff}$ as a function of $\sigma_{\rhostar}$, assuming $\sigma_{\mstar} = 5\%$. The break-even point, where $\sigma_{\teff}=2.0\%$ from \citet{Tayar:2022}, is shown as a vertical red dashed line and corresponds to $\sigma_{\rhostar}\sim 10.3\%$.}
    \label{fig:teff_vs_rhostar}
  \end{center}
\end{figure}

If we assume that $\sigma_{\mstar} = 5\%$, we compute the same $\sigma_{\rstar}$ as in \S \ref{sec:rstar}. Then, we assume $\sigma_{\lstar} = 2.4\%$ and plug those into equation \ref{eq:sigma_teff_frac_lstar} to plot $\sigma_{\teff}$ as a function of $\sigma_{\rhostar}$, in Figure \ref{fig:teff_vs_rhostar}. The break-even point is shown as a vertical red dashed line at $\sigma_{\rhostar} \sim 10.3\%$. That is, measuring the precision in \rhostar \ to better than 10.3\% can improve the precision of the \teff \ to better than the 2\% error floor from \citet{Tayar:2022}.

We note that large systematic errors in \lstar -- far exceeding the 2.4\% \ suggested by \citet{Tayar:2022} -- are possible if we fail to identify visual or bound companions that are blended in the broadband photometry. This would introduce large systematic errors in the SED model, increasing the inferred stellar radius and/or temperature. However, these biases are subject to the same sort of scaling -- a bias of 5\% in \teff \ would require an undetected companion with 20\% of the flux of the primary, which would likely be detected in high-resolution spectroscopy.

\subsubsection{Bolometric flux, distance systematics}
\label{sec:distance}

We can also write the \teff \ uncertainty in terms of the bolometric flux, which is the observed quantity,

\begin{equation}
    \label{eq:fbol}
    \fbol \equiv  \sigmasb \teff^4 \left(\frac{\rstar}{d}\right)^2,
\end{equation}

\noindent where $d$ is the distance to the star. Following our usual procedure, the fractional uncertainty in \teff \ becomes

\begin{equation}
    \label{eq:sigma_teff_frac_fbol}
    \left(\frac{\sigma_{\teff}}{\teff}\right)^2 =
    \left(\frac{\sigma_{d}}{2d}\right)^2 + \left(\frac{\sigma_{\fbol}}{4\fbol}\right)^2 + \left(\frac{\sigma_{\rstar}}{2\rstar}\right)^2
\end{equation}

When the uncertainty in the distance is negligible, equations \ref{eq:sigma_teff_frac_lstar} and \ref{eq:sigma_teff_frac_fbol} are functionally identical. \citet{Tayar:2022} state that the vast majority of planet-hosting stars have negligible distance uncertainties, and they ignore the distance term, not distinguishing between systematics in \fbol \ and systematics in \lstar. 

Indeed, 75\% of planet-hosting stars have fractional distance uncertainties less than the 4.2\% systematic stellar radius floor they found, and so the uncertainty in radius usually dominates the error budget in \lstar \ without an external constraint on \rhostar \ from transits. However, only 40\% of planet hosts have distance uncertainties below 1.7\% -- the smallest systematic uncertainty in the radius we might expect using our method. Therefore, the uncertainty in the parallax -- and its systematic uncertainty -- is an important consideration in general, even for nearby planet-hosting stars. 

We clarify that the systematic floor quoted on \lstar \ from \citet{Tayar:2022} is entirely based on the systematic errors inherent in \fbol, leaving an important additional source of systematic error in the luminosity from the distance.

There has been a wide recognition that \gaia \ DR2 has systematic errors in the measured parallax, with estimates ranging from 30 - 80 $\mu$as that likely depend on magnitude, color, and ecliptic latitude \citep{Lindegren:2018, Stassun:2018, Zinn:2019b}. \gaia \ EDR3/DR3 is better but still has systematics estimated at the ``few tens of $\mu$as'' \citep{Lindegren:2021}. 

Because $d$ is determined from the parallax $\varpi$ 

\begin{equation}
    \label{eq:distance}
    d \equiv \frac{1 \arcsec}{\varpi} \mbox{ pc},
\end{equation}

\noindent we can propagate the uncertainty in $\varpi$, $\sigma_{\varpi}$, to the fractional uncertainty in distance, $\sigma_d/d$, as

\begin{equation}
    \label{eq:sigma_distance}
    \frac{\sigma_d}{d} = \frac{\sigma_{\varpi}}{\varpi}.
\end{equation}

\begin{figure}
  \begin{center}
    \includegraphics[width=3.5in]{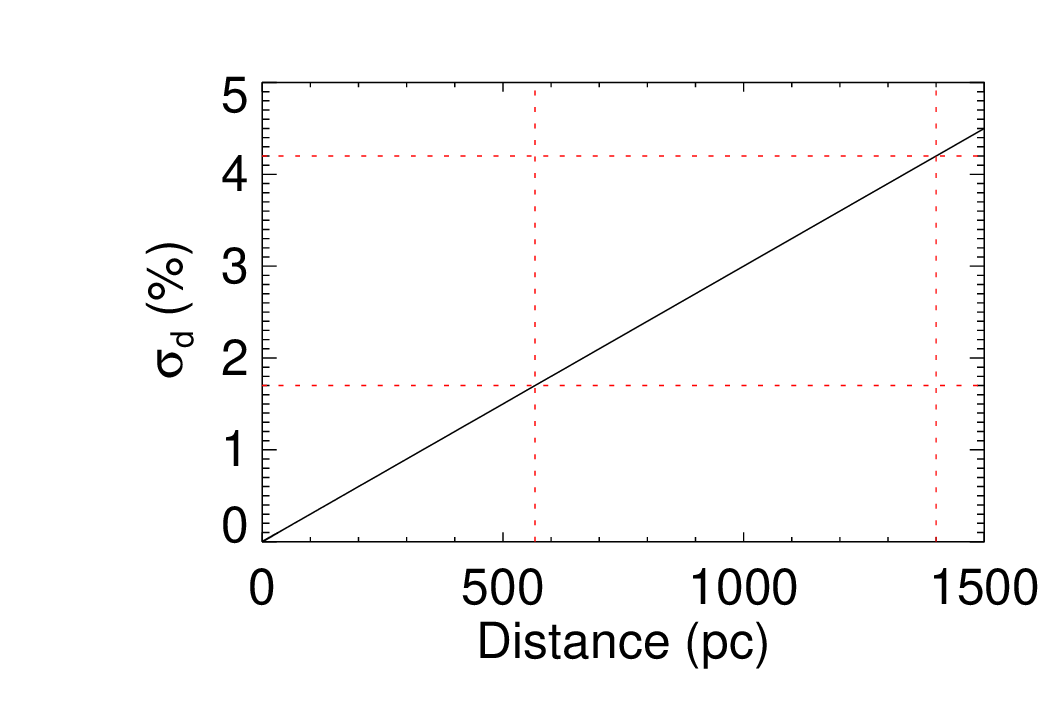}
    \caption{A plot of the level of systematic error in \gaia \ as a function of distance. The black line shows a 30 $\mu$as systematic uncertainty in \gaia \ DR3 estimated by \citet{Lindegren:2021}, which dominates the luminosity systematics at $\sim600$ and $\sim1400$ pc when the \rstar \ uncertainties are at the systematic limits of 1.7\% and 4.2\%, respectively, marked as red dashed lines.}
    \label{fig:distance_errors}
  \end{center}
\end{figure}

We plot Equation \ref{eq:sigma_distance} as a function of distance in Figure \ref{fig:distance_errors}, assuming a systematic floor of 30  $\mu$as from \gaia \ DR3. This systematic floor is the dominant source in the \lstar \ computation for stars beyond 1400 pc when we assume the $\sigma_{\rstar}= 4.2\%$ from \citet{Tayar:2022}, corresponding to 8\% of planet hosts. When we use our systematic floor of 1.7\% using a precise \rhostar, the distance systematic is the dominant source of error in \lstar \ when stars are beyond 567 pc, corresponding to 42\% of planet hosts.

Hence, these systematics cannot, in general, be ignored. Because \gaia \ DR3 has significantly reduced systematic errors, its use is highly recommended over \gaia \ DR2. DR4 is expected to further reduce systematic uncertainties. It is also important to correct for these systematics as best as possible. We note that \exofasttwo \ includes {\tt MKTICSED}, which applies the EDR3/DR3 correction to the parallax described in \citet{Lindegren:2021}, which parameterizes the systematic error as a function of color, magnitude, and ecliptic latitude, but it is unclear what magnitude of systematic error remains.

\subsection{\logg \ from \rhostar \ and \rstar }

Measuring \logg \ from spectra is both imprecise and inaccurate, with systematic error floors of $\sim 0.1$ dex \citep{Torres:2012}.

However, using the same procedure as above, we can propagate errors on \rhostar \ and \rstar \ to \logg \ and achieve a precision more than an order of magnitude better. The fact that the transit can constrain \logg \ has long been appreciated \citep[e.g.][]{Winn:2008}, but it has never been stated in this kind of formalism. 

We start with the definition of the stellar surface gravity,

\begin{equation}
    \label{eq:logg}
    \logg \equiv \log_{10}\left(\frac{G\mstar}{\rstar^2}\right),
\end{equation}

\noindent except we refactor to put it in terms of the directly measured \rhostar \ instead of the systematics-dominated \mstar,

\begin{equation}
    \label{eq:loggrhostar}
    \logg = \log_{10}\left(\frac{4\pi G}{3} \rhostar\rstar\right),
\end{equation}

\noindent and we propagate errors as before,

\begin{equation}
    \label{eq:sigma_logg}
    \sigma_{\logg}^2 = \left(\frac{\partial\logg}{\partial\rhostar}\right)^2 \sigma_{\rhostar}^2 + \left(\frac{\partial\logg}{\partial\rstar}\right)^2 \sigma_{\rstar}^2
\end{equation}

Here the logs already express the \logg \ error in terms of the fractional errors in \rhostar \ and \rstar, so we just evaluate the derivatives and simplify:

\begin{equation}
    \label{eq:sigma_logg_frac_rhostar}
    \sigma_{\logg}^2 = 
    \left(\frac{\sigma_{\rhostar}}{\ln{(10)}\rhostar}\right)^2 + \left(\frac{\sigma_{\rstar}}{\ln{(10)}\rstar}\right)^2
\end{equation}

\begin{figure}
  \begin{center}
    \includegraphics[width=3.5in]{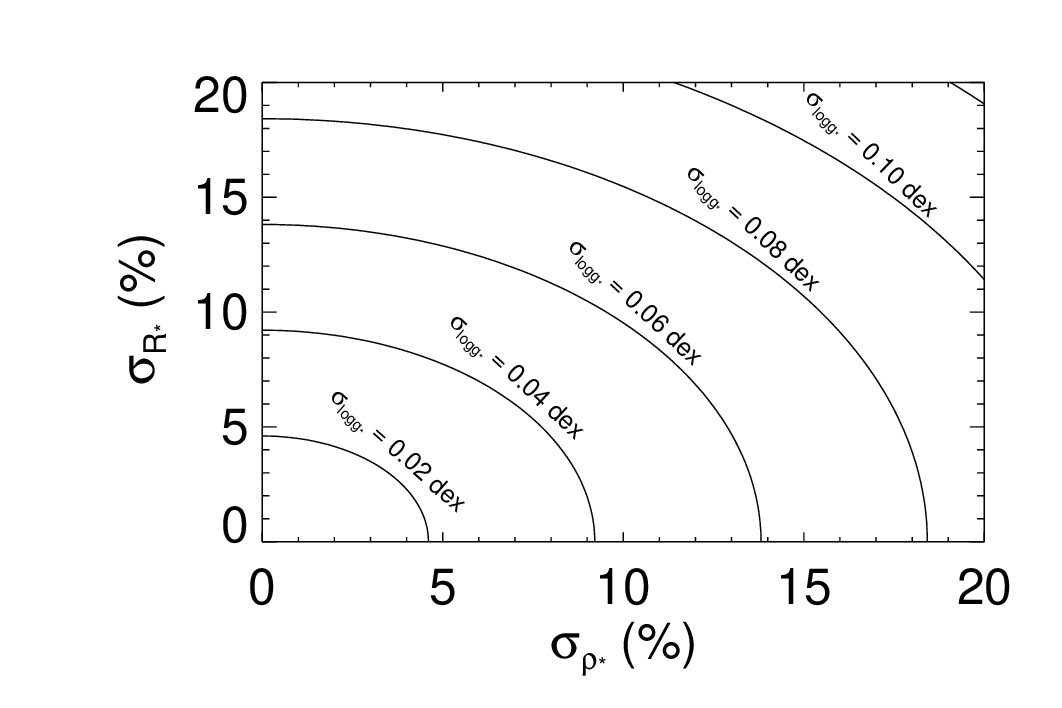}
    \caption{A contour plot of the error in \logg \ as a function of the percent errors in \rhostar \ and \rstar, derived from equation \ref{eq:sigma_logg}. A typical spectroscopic constraint is 0.1 dex. For precise values of \rhostar, we can do more than an order of magnitude better.}
    \label{fig:logg_errors}
  \end{center}
\end{figure}

We show the contour plot of equation \ref{eq:sigma_logg_frac_rhostar} in Figure \ref{fig:logg_errors}, noting that the error in \logg \ is in dex, not percent as for previous, similar plots. A typical spectroscopic constraint on \logg \ is 0.1 dex, which is more than an order of magnitude worse for the best cases where $\sigma_{\rhostar} \sim 1\%$ and $\sigma_{\rstar} \sim 1.7\%$ where we get an uncertainty of  $\sim$ 0.008 dex.

\subsection{\logg \ from \mstar \ and \rstar}
\label{sec:loggfrommr}

In addition, even propagating sensible systematic error floors in \mstar \ and \rstar \ from evolutionary and SED models, we can typically determine a \logg \ that is still more than twice as precise as spectra. That is, in the post-\gaia \ era, we should only rely on a spectroscopic determination of \logg \ in the rare cases where \gaia \ has not measured the distance of a planet host or when the SED cannot be trusted (e.g., due to a blend).

To show this, we repeat the exercise starting from equation \ref{eq:logg}, and again propagate errors,

\begin{equation}
    \label{eq:sigma_logg_mstar}
    \sigma_{\logg}^2 = \left(\frac{\partial\logg}{\partial\mstar}\right)^2 \sigma_{\mstar}^2 + \left(\frac{\partial\logg}{\partial\rstar}\right)^2 \sigma_{\rstar}^2
\end{equation}

\noindent which evaluates to

\begin{equation}
    \label{eq:sigma_logg_mr}
    \sigma_{\logg}^2 = 
    \left(\frac{\sigma_{\mstar}}{\ln{(10)}\mstar}\right)^2 + 
    \left(\frac{2\sigma_{\rstar}}{\ln{(10)}\rstar}\right)^2
\end{equation}

\begin{figure}
  \begin{center}
    \includegraphics[width=3.5in]{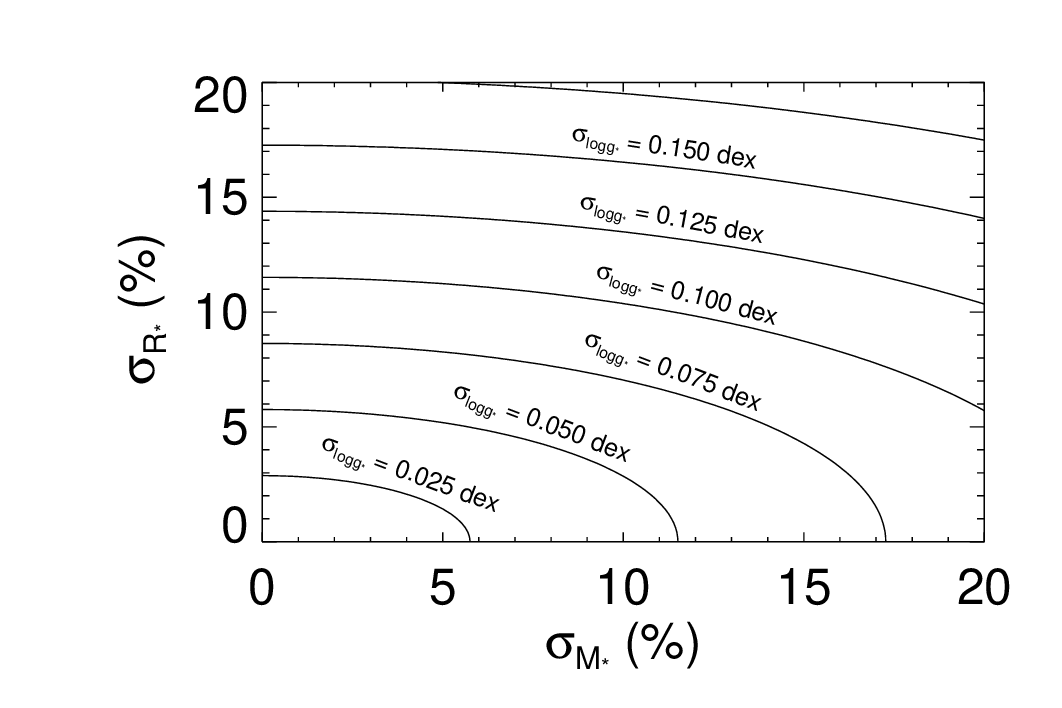}
    \caption{A contour plot of the error in \logg \ as a function of the percent errors in \mstar \ and \rstar, derived from equation \ref{eq:sigma_logg_mr}. A typical spectroscopic constraint of 0.1 dex corresponds to 2\% error for a solar-type star. When we get systematic-error-dominated values for \mstar \ and \rstar \ (without a transit density), we still do about 2x better than spectroscopy.}
    \label{fig:logg_errors_mstar}
  \end{center}
\end{figure}

\noindent We show the contour plot of equation \ref{eq:sigma_logg_mr} in Figure \ref{fig:logg_errors_mstar}. We can see that for the typical exoplanet host star, which is systematics dominated ($\sigma_{\mstar} \sim 5\%$ and $\sigma_{\rstar} \sim 4.2\%$), we get a \logg \ precision of $\sim0.042$ dex -- more than a factor of two better than spectroscopy.

\begin{figure}
  \begin{center}
    \includegraphics[width=3.5in]{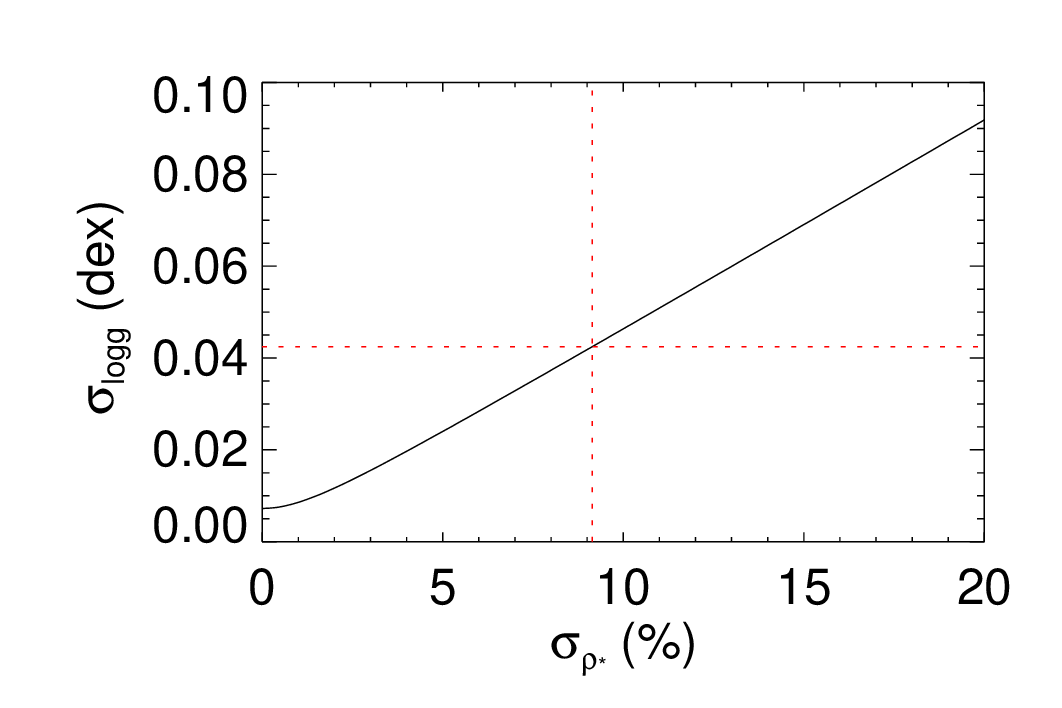}
    \caption{A plot of $\sigma_{\logg}$ as a function of $\sigma_{\rhostar}$, assuming $\sigma_{\mstar} = 5\%$. The break-even point of $\logg=0.042$ dex, where $\sigma_{\mstar}=5\%$ and from \citet{Tayar:2022}, is shown as a vertical red dashed line and corresponds to $\sigma_{\rhostar}\sim 9\%$.}
    \label{fig:logg_vs_rhostar}
  \end{center}
\end{figure}

If we assume that our $\sigma_{\mstar} = 5\%$, we compute the same $\sigma_{\rstar}$ as in \S \ref{sec:rstar}. Then, we plug those into equation \ref{eq:sigma_logg_frac_rhostar} to plot $\sigma_{\logg}$ as a function of $\sigma_{\rhostar}$, in Figure \ref{fig:logg_vs_rhostar}. The break-even point is shown as a vertical red dashed line at $\sigma_{\rhostar} \sim 9\%$. That is, measuring \rhostar \ to better than 9\% can improve the precision of the \logg \ to better than the 0.042 dex derived from the floors in \citet{Tayar:2022}.

\subsection{\mstar \ from \logg \ and \rstar}

Another proposed avenue to get empirical masses for the star when the host does not have a transiting planet is to derive \mstar \ from a \logg \ obtained from something like Flicker \citep{Bastien:2013,Bastien:2016} or asteroseismology plus \rstar \ from an SED. If we propagate the fractional errors in \gstar\  and \rstar\ to \mstar,

\begin{equation}\label{eq:mstar_gstar}
\left(\frac{\sigma_{\mstar}}{\mstar}\right)^2 = \left(\frac{2\sigma_{\rstar}}{\rstar}\right)^2 + \left(\frac{\sigma_{\gstar}}{\gstar}\right)^2,
\end{equation}

\noindent we see that this scaling is more favorable than when starting from \rhostar, but still requires a precise \gstar \ to be competitive with systematic floors when using stellar models, and the fractional uncertainty in \rstar \ is still doubled as it propagates to \mstar.

We note that deriving \mstar \ from a spectroscopic \logg, while possible, is unlikely to be a competitive approach since the systematic uncertainty in a spectroscopic \logg \ is 0.1 dex \citep{Torres:2012}.

\section{Systematic errors in \rhostar}
\label{sec:systematics}

Statistical errors often dominate, and when they do, they are effortlessly propagated throughout the global model. However, we must ensure that our method does not introduce a new systematic error that is large compared to the systematic error in \mstar. Therefore, we must ensure that the systematic errors on \rhostar \ from any source are below 2\%, at which point the total model systematics are dominated by the systematic error already introduced by \mstar.

The derivation of \rhostar \ from transits is straightforward and has been done in many places \citep[e.g.][]{Winn:2010}. For context, we repeat it here. Starting with Kepler's law and the known planetary period, P, 

\begin{equation}
    \label{eq:kepler}
    P^2 = \frac{4\pi^2a^3}{G\left(\mstar + \mplanet\right)},
\end{equation}

\noindent we refactor in terms of \rhostar \ and solve,

\begin{equation}
    \label{eq:rhostararmp}
    \rhostar  = \frac{3\pi}{GP^2}\left(\frac{a}{\rstar}\right)^3 \frac{\mstar}{\mstar+\mplanet}.
\end{equation}

\noindent If we wish, we can refactor equation \ref{eq:rhostararmp} in terms of \rplanet \ and \rhoplanet, which makes the negligible planetary term more obvious:

\begin{equation}
    \label{eq:rhostarar}
    \rhostar = \frac{3\pi}{GP^2}\left(\frac{a}{\rstar}\right)^3 - \left(\frac{\rplanet}{\rstar}\right)^3\rhoplanet.
\end{equation}

Figure \ref{fig:rstar_vs_rhostar} shows that the 5\% systematic error in \mstar \ dominates as long as $\sigma_{\rhostar} \lesssim 2\%$, so here we enumerate potential sources of systematic error to let the reader understand when systematics in \rhostar \ might be the dominant consideration. When the sources of systematic error are well below that floor, no matter what statistical precision we achieve in \rhostar, we can trust the derived uncertainties in \rstar \ and \teff.

If the companion mass is less than 20 \mj \ for a solar-type star, ignoring the planetary mass entirely contributes less than 2\% to the \rhostar, so we drop that term moving forward, and the fractional uncertainty in \rhostar \ becomes

\begin{equation}
    \label{eq:sigma_rhostarar_frac}
    \begin{split}
        \left(\frac{\sigma_{\rhostar}}{\rhostar}\right)^2 = & \left(\frac{\sigma_P}{P}\right)^2 + 
        \left(\frac{3\sigma_{a/\rstar}}{a/\rstar}\right)^2.
    \end{split}
\end{equation}

\subsection{$a/\rstar$}

For the majority of systems, the most problematic component in computing $\rhostar$ is $a/\rstar$. The constraint comes down to the signal-to-noise and our ability to resolve the ingress and egress of the transit, and it is strongly degenerate with the planet's impact parameter. Not only is the measurement less straightforward, but its percent uncertainty is magnified by 3 when propagating to \rhostar, as we see in equation \ref{eq:sigma_rhostarar_frac}.

\subsubsection{Eccentricity}
\label{sec:eccentricity}
First, $a/\rstar$ is not the observable; the transit duration is. For nongrazing, circular orbits, the transit duration translates to a direct constraint on $a/\rstar$, but for eccentric orbits, \citet{Winn:2010} showed that the observable is better approximated by 

\begin{equation}
    \label{eq:arobserved}
    \frac{a}{R_*}\frac{\sqrt{1-e^2}}{1+e\sin{\omega_*}}.
\end{equation}

\noindent This means we must also know or assume the planetary eccentricity and argument of periastron independently from the light curve. \citet[Equation 79 and Figure 10,][]{Stevens:2018} show that, for an eccentricity known to better than $\sim1.5\%$ -- and more lax for smaller eccentricities -- the uncertainty in $e$ contributes negligibly to the uncertainty in \rhostar. We note that \citet{Stevens:2018} assume the covariance between eccentricity and \rhostar \ is negligible, which is true when the eccentricity is measured independently. This is not true when the eccentricity is derived from the light curve itself, but in that case the light curve's power to determine an independent \rhostar \ is limited.

Planets in very short periods can be assumed to be tidally circularized \citep{Adams:2006}. Given the small impact at low eccentricities, even if the planet is not strictly circularized, the error introduced to \rhostar \ is indeed negligible.

For other systems, we must rely on RVs, and we inherit the systematic errors of the spectrograph. The impact on the inferred eccentricity depends heavily on the spectrograph and the planet. In many cases, particularly for hot Jupiters most amenable to measuring \rhostar, the statistical error dominates the systematic error, but for systems where the RV semiamplitude is comparable to the instrumental precision, the systematic uncertainty may dominate.

Or, we can rely on the combination of the timing and duration of both the primary and secondary transit, inheriting the systematic errors of the photometric instrument and the clock. This typically yields extremely precise measurements of eccentricity, well below 1.5\%.

In the future, we may be able to use \gaia \ DR4 to determine the eccentricity for a handful of (nearby, long-period) transiting systems, inheriting the systematic errors on its astrometry.

Ultimately, we need to be mindful of systematic error sources when the eccentricity uncertainty exceeds $\sim 1.5\%$ -- which is often, and is likely to limit the number of stars where we can do such measurements. A campaign to measure precise eccentricities through secondary eclipse timing may dramatically broaden the number of stars where this technique is practical.

\subsubsection{Grazing transits}

When the transit is grazing, significant degeneracies are introduced between the duration (i.e., $a/\rstar$), inclination, and planetary radius. Because of that degeneracy, it is unlikely that grazing transits will provide a sufficient constraint on $a/\rstar$ ($\sigma_{\rhostar} \lesssim 10\%$) to improve the stellar parameters.

\subsubsection{Blending and starspots}

Blending from sources of light other than the star dilutes the transit light curve, biasing \rhostar, as discussed in detail by \citet{Kipping:2014}. Because blending only makes the observed transit depth smaller than it is, it can only underestimate the light-curve-derived stellar density. Equation 9 in \citet{Kipping:2014} computes the bias on \rhostar \ as a function of the observed
$p=\rplanet/\rstar$, 
the observed impact parameter $b$, and the blend fraction $\mathcal{B}$, reproduced here:

\begin{equation}   
\left(\frac{\rho_{*,\obs}}{\rho_{*,\tru}}\right) = \mathcal{B}^{-3/4} \left(\frac{(1+\sqrt{\mathcal{B}} p)^2-b^2}{(1+p)^2-b^2}\right)^{3/2}.
\label{eq:blend}
\end{equation}

We want to know the minimum contaminant, $\mathcal{B}$, that can cause a 2\% error in the stellar density, or $\left(\frac{\rho_{*,\obs}}{\rho_{*,\tru}}\right) = 0.98$. That error is technically unbounded as the numerator approaches zero, or when $b=1+p$. However, as we discussed in the previous section, grazing planets are already problematic and should not be used to infer the stellar density, so we will only consider nongrazing ($b<1-p$) planets.

\begin{figure}
  \begin{center}
    \includegraphics[width=3.5in]{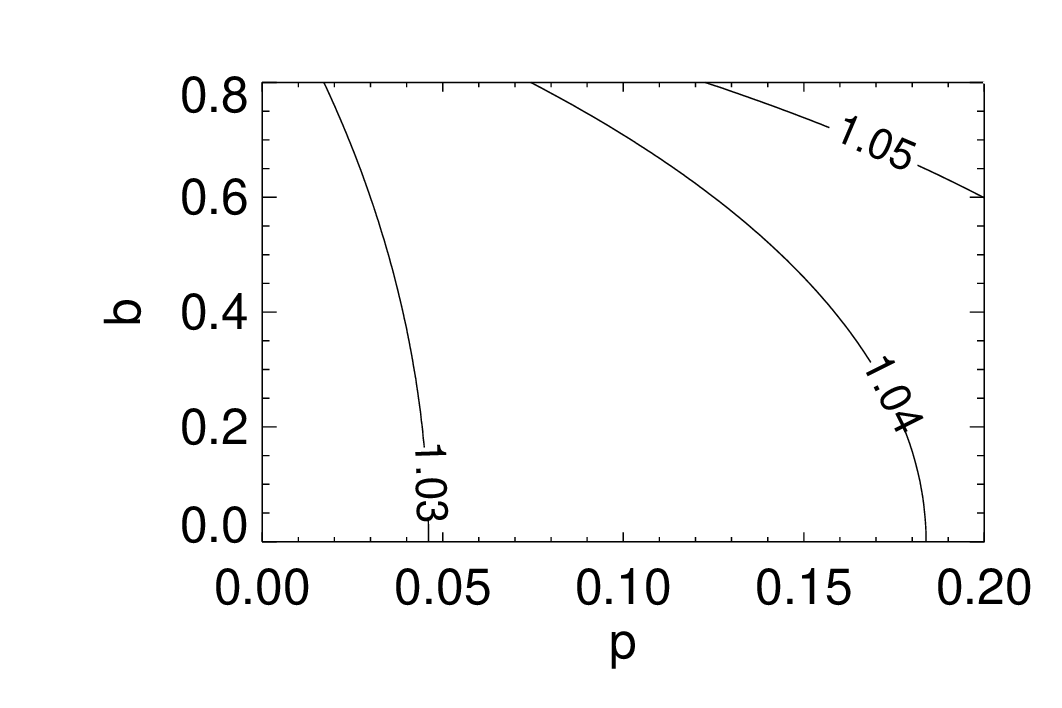}
    \caption{A contour plot showing the maximum allowed unaccounted blend in the transit light curve as a function of impact parameter $b$ and planet-to-star radius ratio $p$ such that its systematic impact \rhostar \ is below the 2\% systematic impact from \mstar. That is, the leftmost contour shows the family of planets whose systematics on the lightcurve-determined \rhostar \ would be dominated by blends larger than 3\% of the host star flux. Stated differently, as long as the unaccounted flux was less than 3\%, the largest source of systematic error in the model would come from \mstar.}
    \label{fig:blend}
  \end{center}
\end{figure}

Figure \ref{fig:blend} shows the maximum allowed blending as a function of $b$ and $p$ such that the systematic error contribution from blending is less than the systematic error contribution from \mstar. For a typical hot Jupiter (p=0.1,b=0.5), this means that the maximum allowed flux from an unseen blended companion is 3.5\%, which is an important consideration.

Starspots have a similar impact to blends, though they may make the star dimmer or brighter than its mean. For spot modulations with a similar amplitude to that above, be mindful of its systematic impact on \rhostar. A more detailed discussion may be found in \citet{Kipping:2014}.

\subsubsection{Transit Timing Variations, Transit Duration Variations, and long integration times}

Transit timing variations (TTVs), transit duration variations (TDVs), and long integration times all smear out the observed transit and systematically bias the inferred stellar density. In general, when TTVs are detectable, they should be accounted for, meaning that only TTVs that are undetectable will bias the inferred stellar density. \citet{Kipping:2014} quantifies this dependence as

\begin{equation}
\sigma_{\rhostar} \lesssim 7.5 \frac{G^{1/3}\rhostar^{1/3}}{p P^{1/3}} \frac{\sigma_{\tau}}{N^{1/4}},
\label{eqn:ttv}
\end{equation}

\noindent which shows that the impact of not accounting for a 1-minute TTV, in the worst case of a central crossing transit, is about 10\% for a Jupiter-sized planet -- nominally five times our floor. For small planets, the impact is worse. Confirmed TTVs for the hot Jupiters most amenable to the best \rhostar \ precision are rare, but this is an important caveat to consider. \citet{Kipping:2014} also shows that the systematic impact of TDVs on \rhostar \ is similar to the impact from TTVs, but less important in practice owing to their relative rarity. When undetectable TTVs are a concern, we could fit for them and would naturally propagate the timing uncertainty into the \rhostar \ from the light curve, though this would necessarily reduce the resultant precision of \rhostar.

In general, long integration times smear out the transit in a way that is similar to TTVs. Fortunately, the exposure times are well-known and that smearing is easily modeled (e.g., in \exofasttwo, for each light curve, the user may specify the exposure times and how many model points to interpolate). However,  failing to do so introduces important systematic errors similar to neglected TTVs with an amplitude of the exposure time. \citet{Price:2014} discuss this impact in detail, but for TESS and Kepler, that is typically well in excess of our target 2\% floor.

\subsubsection{Limb darkening}

An a priori constraint on the limb darkening is often derived from stellar atmospheric models. Theoretical limb-darkening coefficients can differ dramatically from empirical measurements, especially for nonsolar ($\teff \lesssim 5000$ K or $\teff \gtrsim 7000$ K) stars \citep{Patel:2022}. In addition, the common choice of a quadratic limb-darkening law is not, in detail, correct. As both $a/\rstar$ and the limb darkening depend on the shape of the transit, errors in the limb darkening may bias the inferred values of $a/\rstar$.

With sufficiently precise light curves, we can measure the limb darkening directly and remove the reliance on the stellar atmospheric models. Even in simulated cases where the theoretical limb darkening differs from the actual limb darkening by 0.1 in each quadratic term (nominally twice the assumed model uncertainty), its impact on the stellar parameters is negligible. Still, we recommend that when the light curve is sufficiently precise to directly measure the limb darkening, theoretical limb-darkening tables \citet[e.g.,][]{Claret:2011} should not be used to further constrain them. When using \exofasttwo in particular, if the reported precision of the limb-darkening parameters is smaller than the 0.05 systematic error assumed in the tables, the tables should not be used.

In addition, the choice of the limb-darkening law can still bias the fit. \exofasttwo \ only implements the quadratic limb-darkening law, introducing a systematic error floor. However, \citet{Mandel:2002} showed the the error introduced here is typically below the noise floor, and thus negligible as it propagates to \rhostar. 

\subsubsection{Non-Keplerian Motion}

The foundation of our derivation of \rhostar \ is Kepler's law, but the presence of additional bodies and tidal forces means that nothing follows Kepler's law to infinite precision. For a system with TTVs, $a/\rstar$ changes over time, but surely that has no impact on the stellar density. \exofasttwo \ assumes Keplerian orbits, but computational time is the only reason we hesitate to implement an N-body code to compute the planetary orbits. Given that the transit duration and stellar density are never explicitly defined in the transit model, it is likely that an accurate computation of any non-Keplerian motion would provide a similar constraint on the stellar density, but a detailed investigation of this is beyond the scope of this paper.

\subsection{Period}

The planetary period is directly measured from the frequency of transits, leveraged with long baselines between transits, often resulting in part-per-billion precision in the planetary period. However, it is worth noting that we universally introduce a systematic error in the observed period that is statistically significant in many systems today. Because stars are moving with respect to the solar system barycenter frame, there is a light-travel time effect that changes the observed frequency of transits by the systemic velocity, $\gamma$, divided by the speed of light, $c$

\begin{equation}
    \label{eq:gammac}
    \Delta P = \gamma/c.
\end{equation}

But the reported planetary period is universally given in the solar system barycenter frame. Given a typical systemic velocity of $\sim10$ km s$^{-1}$, this $\sim$30 ppm effect amounts to about 30 s in a 10-day period, which is easily measurable today for the vast majority of transiting systems. Even \exofasttwo, which does transform the observed times to the target frame before computing the model, ignores this effect because the RV is often measured from a reference spectrum taken at an earlier time, and so the absolute systemic velocity is often unknown. In addition, reporting the true period in the target's barycentric frame would lead to confusion in propagating the ephemerides that are practically important.

However, the impact of this error on \rhostar \ (or any observable we care about) is dwarfed by other errors. The 30 ppm effect is 5000 times lower than our threshold, so we safely ignore it. 

\section{Implementation}
\label{sec:implementation}
Despite the simplicity of the argument, in many cases a fundamentally new approach must be developed to take advantage of it. We can no longer simply interpolate an evolutionary grid to find the stellar parameters, as is commonly done. Nor can we simply separate the stellar and planetary model, as is also often done. The additional stellar density constraint overconstrains the evolutionary model grids, requiring optimization of competing constraints while simultaneously respecting the systematic error floors inherent in the evolutionary and atmospheric models. In addition, the constraints are often correlated in important ways, and those covariances must be known and applied with care if iterating between the stellar and planetary models to improve the precision of both. It is not enough to apply Gaussian, uncorrelated priors with each iteration.

As far as we are aware, \exofasttwo \ is unique in this regard -- among private and public exoplanet modeling codes. The link between the stellar density and the transit photometry has been at the heart of \exofast \ since its inception \citep{Eastman:2013}, and the link between \teff, \rstar, and \lstar \ has been coded within \exofasttwo \ since SED fitting was added in 2017 January \citep{Eastman:2019}. Despite only deeply understanding the mechanism now, \exofasttwo \ has long respected these relations and has been capable of using the transit-derived density to determine stellar parameters that are less dependent on the systematic floors of evolutionary models.

However, until \gaia \ DR2 in 2018 April, we could not always measure sufficiently precise luminosities, and up until 2020 October, we ignored systematic errors in the SED model, which resulted in many fits with underestimated uncertainties. In July 2022, another update now allows users to specify their own systematic error floors on the stellar evolutionary models so that they may more accurately reflect those found by \citet{Tayar:2022} rather than use the default ad hoc systematic error floors as a function of stellar mass described in \citet{Eastman:2019} and summarized in equation \ref{eq:mistuncertainty}:

\begin{equation}
    \label{eq:mistuncertainty}
    \sigma = 0.03 - 0.025\log{\mstar} + 0.045\left(\log{\mstar}\right)^2.
\end{equation}

We warn the user that the sample used by \citet{Tayar:2022} consisted of near solar-type stars, and for such stars \citet{Tayar:2022} showed that the default $\sim$3\% errors \exofasttwo \ uses are likely slight underestimates of the systematic floors. However, for lower-mass stars, the systematic errors are likely much larger than the sample \citet{Tayar:2022} explored, and our default ad hoc value of $\sim$10\% is likely more appropriate. In addition, the detailed results from \citet{Tayar:2022} were highly system dependent, and our blanket values of the systematic floor are simplified for the sake of presentation.

Finally, while the true nature of systematic errors is still poorly understood, we presume that all theoretical models share similar systematics, and so any combination of stellar theoretical models should not drive the uncertainties below the floors described in \citet{Tayar:2022}. When using multiple theoretical models (e.g., MIST and SED) with separate floors on the same parameters, it may be necessary to further inflate the individual systematic errors to achieve the desired total systematic floor in a star-only fit. Checking that the floors are as desired with a star-only fit is a good standard practice.

\section{WASP-4\lowercase{b}}
\label{sec:wasp4}

While this analytic derivation is helpful for understanding where the information comes from and why, we assume that errors are Gaussian and uncorrelated, which is not strictly true. In this section, we model WASP-4b using \exofasttwo \ \citep{Eastman:2019} to confirm and validate our analytic formulae. A Markov Chain Monte Carlo code like \exofasttwo \ does not assume the errors are Gaussian or uncorrelated, and so we can check that our analytic assumptions are reasonable by fitting a real-world system with a variety of constraints to check for such correlations and to see whether they re-create our uncorrelated expectations. 

WASP-4b is a planet in a short period (1.34 days) that we can reasonably assume is tidally circularized, and so we know the eccentricity exactly, improving the precision in \rhostar \ (see \S \ref{sec:eccentricity} and \S \ref{sec:tidalcirc}). In addition, it has been observed in three TESS sectors at 2-minute cadence (which can be found in MAST: \dataset[https://doi.org/10.17909/t9-nmc8-f686]{https://doi.org/10.17909/t9-nmc8-f686}), with a very long baseline between the TESS observations and the eight discovery light curves in 2007 \citep{Wilson:2008, Gillon:2009, Winn:2009}. Many of those discovery light curves were observed in Sloan z' band and the transit is nearly edge-on, which minimizes the covariance between density, impact parameter, and limb darkening. Finally, the transit is very deep (2.4\%). All of these combine to enable us to measure the stellar density of WASP-4 to extreme precision. 

A detailed exploration of what contributes to the statistical precision of \rhostar \ and an exhaustive search for the best candidate(s) is beyond the scope of this paper, but for the reasons above, WASP-4 is likely among the best-suited exoplanet hosts for measuring \rhostar. At any rate, for using \rhostar \ to measure \rstar \ and \teff, there are diminishing returns beyond a \rhostar \ precision of $\sim2\%$ because the systematic floor in \mstar \ begins to dominate.

Because of the way the evolutionary model is implemented with \exofasttwo, we can only impose error floors in derived quantities like age, \rstar, \feh, and \teff, not the grid parameters \mstar, \initfeh, and the equal evolutionary phase \citep[EEP; see][]{Dotter:2016}. Further, it was unclear to us how the systematic floors from the evolutionary models might combine with the systematic floors from the atmospheric models within \exofasttwo. We presume that both are limited by our understanding of the underlying stellar astrophysics, and so they should not combine as independent constraints. Instead, the combined MIST+SED models should still be limited by these same systematic floors: 2.4\% in \fbol, 4.2\% \ in \rstar, 5\% in \mstar, 2.0\% in \teff, and 0.08 dex in \feh. However, this is complicated by the fact that we cannot tune these final floors directly, and the physical relation between these parameters often means that we cannot respect all floors exactly and simultaneously.

We began by doing a preliminary fit of only the WASP-4 host star including an SED fit of \gaia, 2MASS, and WISE broadband photometry; a MIST stellar evolutionary model \citep{Paxton:2011,Paxton:2013,Paxton:2015,Dotter:2016,Choi:2016}; priors on $\feh=-0.03 \pm 0.09$ \citep{Gillon:2009}; parallax=$3.797 \pm 0.061$ mas \citep{Gaia:2018}; and an upper limit on the V-band extinction of 0.04278 mag based on galactic dust maps \citep{Schlegel:1998, Schlafly:2011}. While a spectroscopic prior was available for \teff, we chose not to use it, as the $\sim2\%$ systematic uncertainty expected is much higher than the uncertainty we expect from our method described in \S \ref{sec:tefffromlstar}. For reference, \citet{Wilson:2008} found $\teff=5500 \pm 150$ K using CORALIE, and \citet{Gillon:2009} found $\teff = 5470 \pm 130$ K using IRFM, which is in good agreement ($0.4\sigma$) with our final recommended value of $5419^{+65}_{-63}$ K from the MIST + transit + SED fit. 

We first fit a MIST+SED model with floors in the evolutionary model \rstar \ of 4.2\%, \teff \ of 2.0\%, and \feh \ of 0.08 dex, and SED model floors of 2.4\% in \fbol, 2.0\% in \teff, and 0.08 dex in \feh. However, the combined constraint was lower than our model floors should be trusted. Hence, we inflated the \rstar and \teff \ systematic floors by $\sqrt{2}$ and refit. The model floors were still not exactly as desired because we cannot match them all at once owing to their influence on one another. They were close, but the best way to reconcile these competing constraints is unclear. In the MIST+SED column of table \ref{tab:wasp4}, we see that our final constraints are close to our desired floors: 3.2\% in \fbol, 3.8\% \ in \rstar, 5.5\% in \mstar, 2.2\% in \teff, and 0.084 dex in \feh.

It would be best to do a thorough investigation that explores systematic differences between models similar to \citet{Tayar:2022} or \citet{Duck:2022} for each modeled system and set these floors accordingly, but this is a major effort and likely impractical for all systems.

Next, we performed eight different fits of the WASP-4 system with all combinations of with and without the SED model, the MIST model, and the transit model, including no model constraints, labeled ``None,'' showing just our prior constraints. Each fit was constrained with the same wide, uniform priors summarized in Table \ref{tab:wasp4priors}, equal to five times the 68\% confidence interval of the preliminary MIST+SED fit described above. These are wide enough not to appreciably influence fits that were reasonably well constrained, but narrow enough to allow the fits to mix in the absence of any external constraints, see the impact of our chosen stepping parameters, and ensure that the only things that changed between fits were the models used to constrain them. In the case where we do not fit the SED, MIST, or transit model, the posteriors are equal to these priors.  All fits had the same systematic error floors imposed where applicable. 

\begin{deluxetable}{llc}
\startlongtable
\tablecaption{Priors imposed on all WASP-4b fits}
\tablehead{\colhead{Parameter} & \colhead{Units} & \colhead{Prior}}
\startdata
$M_*$          \dotfill& Mass (\msun)                 \dotfill& $\mathcal{U}[0.694,1.104]$\\
$R_*$          \dotfill& Radius (\rsun)               \dotfill& $\mathcal{U}[0.83,0.97]$\\
$T_{\rm eff}$  \dotfill& Effective temperature (K)    \dotfill& $\mathcal{U}[5120,5735]$\\
$[{\rm Fe/H}]$ \dotfill& Metallicity (dex)            \dotfill& $\mathcal{G}[-0.03,0.09]$\\
$A_V$          \dotfill& V-band extinction (mag)      \dotfill& $\mathcal{U}[0,0.04278]$\\
$\varpi$       \dotfill& Parallax (mas)               \dotfill& $\mathcal{G}[3.7965,0.0608]$\\
$e$            \dotfill& Eccentricity                 \dotfill& 0 (fixed)\\
$\omega_*$     \dotfill& Argument of periastron (deg) \dotfill& 90 (fixed)
\enddata
\label{tab:wasp4priors}
\tablenotetext{}{}
\end{deluxetable}

For all fits including transits, we included the 14 discovery RVs from CORALIE \citep{Wilson:2008}; eight early, complete light curves \citep{Wilson:2008, Gillon:2009, Winn:2009} and the flattened, 2-minute SPOC TESS light curves from sectors 2, 28, and 29. The transit data spanned  13 yr and 3545 epochs. We disabled the limb-darkening table look-up from \citet{Claret:2011} to avoid introducing any systematic errors \citep{Patel:2022} and fit the quadratic limb-darkening parameters in each band directly. We assumed that the orbit was circular.

\begin{figure*}
  \begin{center}
    \includegraphics[width=10in, trim={3cm 0cm 1cm 0cm}, clip]{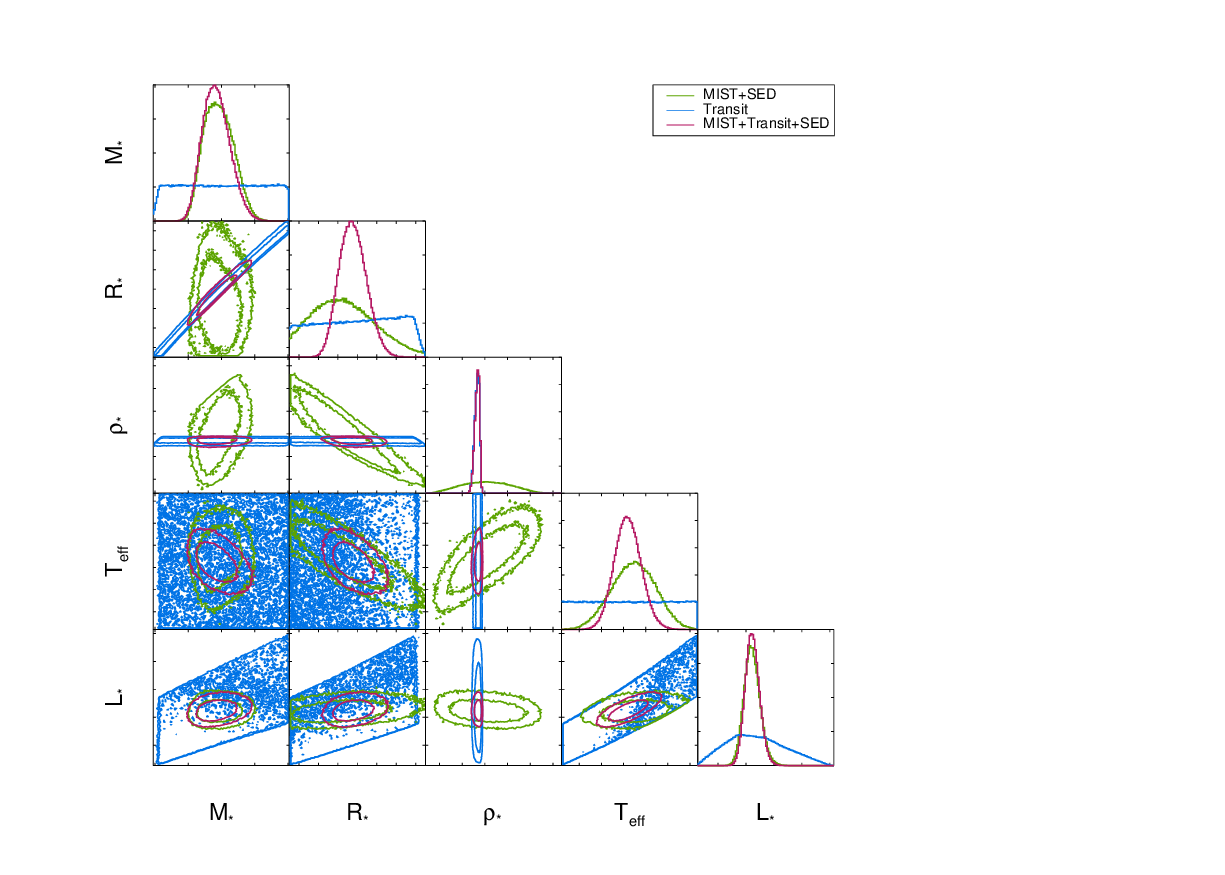}
    \caption{A corner plot of the WASP-4 stellar parameters. The contours show the 68\% and 95\% confidence intervals for the MIST+SED fit (green), the transit-only fit (blue), and the MIST+transit+SED fit (magenta). From this, we can see just how the addition of the transit constrains the stellar density and significantly improves the precision of \rstar \ and \teff \ -- to well beyond the systematic floors imposed on the SED and MIST evolutionary models. The areas under each curve in the histograms is proportional to likelihood and are normalized to have the same area between each of the fits for a given parameter. For \rhostar \ in particular, we can see the huge impact the transit has on shifting the probability to a narrow, high peak.}
    \label{fig:wasp4covar}
  \end{center}
\end{figure*}

In figure \ref{fig:wasp4covar}, we show the corner plot of the stellar parameters for the three most relevant fits -- the MIST+SED, transit-only, and MIST+transit+SED. As expected, we see that \rhostar \ uncertainty is dramatically reduced with the transit, and due to its covariance with \rstar \ and \teff, their uncertainties are also significantly reduced. We also see that the combination of MIST, SED, and the transit is somewhat more complex than our mathematical assumption that the errors are Gaussian and uncorrelated. The slight covariance between \mstar \ and \rstar \ in the MIST+SED fit means that when we add the transit, we also slightly improve the constraint on \mstar \ ($\sim10$\%). 

As can be seen in the last column of Table \ref{tab:wasp4}, using MIST, the SED, and the transit model allows us to measure the stellar density to 1.2\%. Despite the models using the floors above, we are able to infer \mstar \ to 4.8\%, \rstar \ to 1.6\%, and \teff \ to 1.1\%, in line with our analytic expectations given such a precise \rhostar. We note that here we do not achieve the 0.9\% precision expected from Figure \ref{fig:teff_errors} because the uncertainty in the distance is not negligible and we do not reach the floor in \fbol, likely because WASP-4 is a relatively faint planet host (V=12.5). With the $\sigma_d=1.5\%$, $\sigma_{\fbol}=2.9\%$, and $\sigma_{\rstar}=1.6\%$ we achieve,  equation \ref{eq:sigma_teff_frac_fbol} predicts a $\sigma_{\teff}=1.3\%$, in good agreement with our measurement of 1.1\%.

The improvement of the planetary parameters, summarized in Tables \ref{tab:wasp4b} and \ref{tab:wasp4b_notransit}, is also significant. However, we must be careful because \exofasttwo \ cannot sever the connection between the transit model and the stellar density. Table \ref{tab:wasp4b} shows all the fits that include the transit and so are direct outputs from \exofasttwo. Table \ref{tab:wasp4b_notransit} shows the rederived planetary parameters using the star-only values in the corresponding column of Table \ref{tab:wasp4} combined with the transit-observables taken from the transit-only fit -- thus re-creating methods that model the star and planet separately.

These transit-observables span all columns in Table \ref{tab:wasp4b_notransit} and agree with each other to at least $0.1\sigma$ in all fits using a transit in Table \ref{tab:wasp4b}. Thus, comparing the MIST+Transit+SED column in Table \ref{tab:wasp4b} with the MIST+SED column in Table \ref{tab:wasp4b_notransit} shows the improvement in the planetary parameters that is achievable when we account for realistic systematic floors in the stellar models, we have a strong constraint on stellar density, and we model the star and planet simultaneously. 

Most importantly, the precision in the planet's radius, density, surface gravity, semi-major axis, and incident flux improves by about a factor of two. The improvement in \teq \ is equally significant, but it is likely that our statistical uncertainties are dominated by the assumptions that there is no albedo and perfect redistribution. However, the incident flux is improved by a similar factor and is an important, fundamental component of the detailed atmospheric modeling necessary to truly understand the equilibrium temperature and habitability more broadly.

\clearpage
\begin{turnpage}

\begin{table*}
	\centering
	\caption{Median values and 68\% confidence interval for the WASP-4 host star with all 8 combinations of using or not using the SED, MIST, or Transit constraint.}
	\label{tab:wasp4}
	\begin{tabular}{llcccccccc}
	\toprule	\colhead{~~~Parameter} & \colhead{Units} & \colhead{None} & \colhead{MIST} & \colhead{SED} & \colhead{MIST+SED} & \colhead{Transit} & \colhead{Transit+SED} & \colhead{MIST+Transit} & \colhead{MIST+Transit+SED} \\
	\toprule
 ~~~~$M_*$\dotfill&Mass (\msun)\dotfill &$0.88^{+0.15}_{-0.13}$&$0.879^{+0.056}_{-0.048}$&$0.88^{+0.15}_{-0.13}$&$0.895^{+0.052}_{-0.046}$&$0.91^{+0.13}_{-0.14}$&$0.88^{+0.13}_{-0.11}$&$0.890^{+0.070}_{-0.057}$&$0.886^{+0.046}_{-0.040}$\\
~~~~$R_*$\dotfill&Radius (\rsun)\dotfill &$0.900^{+0.046}_{-0.047}$&$0.890^{+0.049}_{-0.041}$&$0.890^{+0.044}_{-0.038}$&$0.886^{+0.035}_{-0.032}$&$0.903^{+0.043}_{-0.048}$&$0.893^{+0.041}_{-0.038}$&$0.898^{+0.023}_{-0.020}$&$0.896^{+0.015}_{-0.014}$\\
~~~~$L_*$\dotfill&Luminosity (\lsun)\dotfill &$0.631^{+0.12}_{-0.097}$&$0.63^{+0.12}_{-0.10}$&$0.624^{+0.029}_{-0.028}$&$0.624^{+0.030}_{-0.027}$&$0.631^{+0.12}_{-0.097}$&$0.624^{+0.028}_{-0.026}$&$0.65\pm0.11$&$0.625^{+0.028}_{-0.026}$\\
~~~~$\rho_*$\dotfill&Density (cgs)\dotfill &$1.70^{+0.38}_{-0.31}$&$1.76^{+0.24}_{-0.23}$&$1.74^{+0.37}_{-0.30}$&$1.82^{+0.24}_{-0.22}$&$1.737^{+0.019}_{-0.025}$&$1.737^{+0.019}_{-0.025}$&$1.738^{+0.018}_{-0.025}$&$1.738^{+0.018}_{-0.024}$\\
~~~~$\log{g}$\dotfill&Surface gravity (cgs)\dotfill &$4.473\pm0.074$&$4.483^{+0.039}_{-0.042}$&$4.480^{+0.074}_{-0.073}$&$4.495^{+0.041}_{-0.042}$&$4.484^{+0.020}_{-0.024}$&$4.479^{+0.020}_{-0.019}$&$4.481^{+0.011}_{-0.010}$&$4.4809^{+0.0080}_{-0.0078}$\\
~~~~$T_{\rm eff}$\dotfill&Effective Temp (K)\dotfill &$5430^{+210}_{-200}$&$5440^{+180}_{-190}$&$5440\pm120$&$5450\pm100$&$5420\pm210$&$5430\pm120$&$5470^{+180}_{-200}$&$5419^{+65}_{-63}$\\
~~~~$[{\rm Fe/H}]$\dotfill&Metallicity (dex)\dotfill &$-0.031\pm0.087$&$-0.025\pm0.089$&$0.011^{+0.085}_{-0.083}$&$0.012^{+0.082}_{-0.080}$&$-0.031^{+0.091}_{-0.090}$&$0.013^{+0.082}_{-0.079}$&$-0.021^{+0.089}_{-0.088}$&$0.015^{+0.082}_{-0.078}$\\
~~~~$R_{*,SED}$\dotfill&Radius$^{1}$ (\rsun)\dotfill &--&--&$0.882^{+0.017}_{-0.016}$&$0.882^{+0.017}_{-0.016}$&--&$0.882\pm0.016$&--&$0.883^{+0.016}_{-0.015}$\\
~~~~$F_{Bol}$\dotfill&Bol Flux $\times 10^{10}$ (cgs)\dotfill &--&--&$2.867^{+0.098}_{-0.088}$&$2.870^{+0.095}_{-0.088}$&--&$2.866^{+0.092}_{-0.082}$&--&$2.867^{+0.088}_{-0.082}$\\
~~~~$T_{\rm eff,SED}$\dotfill&Effective Temp$^{1}$ (K)\dotfill &--&--&$5461^{+34}_{-31}$&$5462^{+33}_{-30}$&--&$5461^{+31}_{-29}$&--&$5460^{+30}_{-29}$\\
~~~~$[{\rm Fe/H}]_{SED}$\dotfill&Metallicity (dex)\dotfill &--&--&$0.038^{+0.11}_{-0.092}$&$0.042^{+0.11}_{-0.096}$&--&$0.043^{+0.10}_{-0.089}$&--&$0.043^{+0.11}_{-0.087}$\\
~~~~$A_V$\dotfill&V-band ext (mag)\dotfill &--&--&$0.022^{+0.014}_{-0.015}$&$0.023^{+0.014}_{-0.015}$&--&$0.023^{+0.014}_{-0.015}$&--&$0.023^{+0.014}_{-0.015}$\\
~~~~$\sigma_{SED}$\dotfill&SED error scaling \dotfill &--&--&$1.02^{+0.41}_{-0.24}$&$1.01^{+0.40}_{-0.24}$&--&$0.98^{+0.34}_{-0.22}$&--&$0.97^{+0.33}_{-0.21}$\\
~~~~$\varpi$\dotfill&Parallax (mas)\dotfill &--&--&$3.791\pm0.061$&$3.791\pm0.061$&--&$3.791^{+0.059}_{-0.060}$&--&$3.789^{+0.059}_{-0.060}$\\
~~~~$d$\dotfill&Distance (pc)\dotfill &--&--&$263.8^{+4.3}_{-4.2}$&$263.8^{+4.3}_{-4.2}$&--&$263.8^{+4.2}_{-4.1}$&--&$264.0^{+4.2}_{-4.1}$\\
~~~~$[{\rm Fe/H}]_{0}$\dotfill&Initial Metallicity$^{2}$ \dotfill &--&$-0.00\pm0.11$&--&$0.03\pm0.10$&--&--&$0.00\pm0.11$&$0.033^{+0.10}_{-0.099}$\\
~~~~$Age$\dotfill&Age (Gyr)\dotfill &--&$7.0^{+4.4}_{-3.7}$&--&$6.2^{+4.7}_{-3.9}$&--&--&$6.6^{+4.4}_{-4.0}$&$7.3^{+3.7}_{-3.8}$\\
~~~~$EEP$\dotfill&Equal Evol Phase$^{3}$ \dotfill &--&$351^{+37}_{-18}$&--&$350^{+37}_{-27}$&--&--&$351^{+33}_{-23}$&$356^{+32}_{-22}$\\
  \hline
	\end{tabular}
\tablenotetext{}{See Table 3 in \citet{Eastman:2019} for a detailed description of all parameters}
\tablenotetext{1}{This value ignores the systematic error and is for reference only}
\tablenotetext{2}{The metallicity of the star at birth}
\tablenotetext{3}{Corresponds to static points in a star's evolutionary history. See \S2 in \citet{Dotter:2016}.}
\end{table*}
\end{turnpage}
\clearpage

\clearpage
\newcommand{\rptabfourone}{$1.348^{+0.064}_{-0.071}$}
\newcommand{\mptabfourone}{$1.22\pm0.13$}
\newcommand{\atabfourone}{$0.0230^{+0.0011}_{-0.0012}$}
\newcommand{\teqtabfourone}{$1638^{+64}_{-63}$}
\newcommand{\tcirctabfourone}{$0.00277^{+0.00021}_{-0.00020}$}
\newcommand{\rhoptabfourone}{$0.620^{+0.045}_{-0.042}$}
\newcommand{\loggptabfourone}{$3.222^{+0.022}_{-0.023}$}
\newcommand{\favetabfourone}{$1.64^{+0.27}_{-0.24}$}
\newcommand{\msinitabfourone}{$1.22\pm0.13$}
\newcommand{\periodtabfourone}{$1.338231512\pm0.000000019$}
\newcommand{\tctabfourone}{$2454697.797524\pm0.000024$}
\newcommand{\tttabfourone}{$2454697.797524\pm0.000024$}
\newcommand{\tstabfourone}{$2454698.466639\pm0.000024$}
\newcommand{\tzerotabfourone}{$2455337.472187\pm0.000022$}
\newcommand{\itabfourone}{$88.87^{+0.67}_{-0.51}$}
\newcommand{\btabfourone}{$0.108^{+0.048}_{-0.064}$}
\newcommand{\ktabfourone}{$240\pm12$}
\newcommand{\ptabfourone}{$0.15348^{+0.00047}_{-0.00039}$}
\newcommand{\artabfourone}{$5.481^{+0.020}_{-0.026}$}
\newcommand{\deltatabfourone}{$0.02356^{+0.00014}_{-0.00012}$}
\newcommand{\deltartabfourone}{$0.03042^{+0.00086}_{-0.00081}$}
\newcommand{\deltaztabfourone}{$0.02781^{+0.00049}_{-0.00046}$}
\newcommand{\deltattabfourone}{$0.02867^{+0.00076}_{-0.00072}$}
\newcommand{\tautabfourone}{$0.01220^{+0.00017}_{-0.00013}$}
\newcommand{\tonefourtabfourone}{$0.08992\pm0.00014$}
\newcommand{\tfwhmtabfourone}{$0.07770^{+0.00017}_{-0.00018}$}
\newcommand{\ruonetabfourone}{$0.460\pm0.041$}
\newcommand{\rutwotabfourone}{$0.152^{+0.075}_{-0.076}$}
\newcommand{\zuonetabfourone}{$0.312^{+0.025}_{-0.024}$}
\newcommand{\zutwotabfourone}{$0.219^{+0.055}_{-0.057}$}
\newcommand{\tuonetabfourone}{$0.364^{+0.041}_{-0.040}$}
\newcommand{\tutwotabfourone}{$0.141^{+0.075}_{-0.076}$}
\newcommand{\rptabfourtwo}{$1.335^{+0.062}_{-0.058}$}
\newcommand{\mptabfourtwo}{$1.20^{+0.13}_{-0.11}$}
\newcommand{\atabfourtwo}{$0.02277^{+0.0010}_{-0.00099}$}
\newcommand{\teqtabfourtwo}{$1639^{+35}_{-36}$}
\newcommand{\tcirctabfourtwo}{$0.00279\pm0.00021$}
\newcommand{\rhoptabfourtwo}{$0.624^{+0.042}_{-0.041}$}
\newcommand{\loggptabfourtwo}{$3.222^{+0.021}_{-0.023}$}
\newcommand{\favetabfourtwo}{$1.64^{+0.15}_{-0.14}$}
\newcommand{\msinitabfourtwo}{$1.20^{+0.13}_{-0.11}$}
\newcommand{\periodtabfourtwo}{$1.338231512\pm0.000000019$}
\newcommand{\tctabfourtwo}{$2454697.797525\pm0.000024$}
\newcommand{\tttabfourtwo}{$2454697.797525\pm0.000024$}
\newcommand{\tstabfourtwo}{$2454698.466641\pm0.000024$}
\newcommand{\tzerotabfourtwo}{$2455348.178040\pm0.000022$}
\newcommand{\itabfourtwo}{$88.87^{+0.67}_{-0.51}$}
\newcommand{\btabfourtwo}{$0.109^{+0.048}_{-0.064}$}
\newcommand{\ktabfourtwo}{$240\pm12$}
\newcommand{\ptabfourtwo}{$0.15349^{+0.00047}_{-0.00040}$}
\newcommand{\artabfourtwo}{$5.481^{+0.020}_{-0.026}$}
\newcommand{\deltatabfourtwo}{$0.02356^{+0.00014}_{-0.00012}$}
\newcommand{\deltartabfourtwo}{$0.03044^{+0.00086}_{-0.00082}$}
\newcommand{\deltaztabfourtwo}{$0.02782^{+0.00049}_{-0.00047}$}
\newcommand{\deltattabfourtwo}{$0.02869^{+0.00075}_{-0.00071}$}
\newcommand{\tautabfourtwo}{$0.01220^{+0.00017}_{-0.00013}$}
\newcommand{\tonefourtabfourtwo}{$0.08992\pm0.00014$}
\newcommand{\tfwhmtabfourtwo}{$0.07770^{+0.00017}_{-0.00018}$}
\newcommand{\ruonetabfourtwo}{$0.461^{+0.041}_{-0.042}$}
\newcommand{\rutwotabfourtwo}{$0.151\pm0.076$}
\newcommand{\zuonetabfourtwo}{$0.313\pm0.025$}
\newcommand{\zutwotabfourtwo}{$0.218^{+0.056}_{-0.057}$}
\newcommand{\tuonetabfourtwo}{$0.365\pm0.040$}
\newcommand{\tutwotabfourtwo}{$0.139^{+0.074}_{-0.075}$}
\newcommand{\rptabfourthree}{$1.341^{+0.035}_{-0.030}$}
\newcommand{\mptabfourthree}{$1.210^{+0.086}_{-0.081}$}
\newcommand{\atabfourthree}{$0.02287^{+0.00060}_{-0.00052}$}
\newcommand{\teqtabfourthree}{$1651^{+54}_{-61}$}
\newcommand{\tcirctabfourthree}{$0.00278\pm0.00019$}
\newcommand{\rhoptabfourthree}{$0.621\pm0.036$}
\newcommand{\loggptabfourthree}{$3.222^{+0.021}_{-0.023}$}
\newcommand{\favetabfourthree}{$1.69^{+0.23}_{-0.24}$}
\newcommand{\msinitabfourthree}{$1.210^{+0.086}_{-0.081}$}
\newcommand{\periodtabfourthree}{$1.338231512\pm0.000000019$}
\newcommand{\tctabfourthree}{$2454697.797524^{+0.000024}_{-0.000023}$}
\newcommand{\tttabfourthree}{$2454697.797524^{+0.000024}_{-0.000023}$}
\newcommand{\tstabfourthree}{$2454698.466640^{+0.000024}_{-0.000023}$}
\newcommand{\tzerotabfourthree}{$2455336.133956\pm0.000022$}
\newcommand{\itabfourthree}{$88.89^{+0.66}_{-0.52}$}
\newcommand{\btabfourthree}{$0.106^{+0.049}_{-0.063}$}
\newcommand{\ktabfourthree}{$240\pm12$}
\newcommand{\ptabfourthree}{$0.15348^{+0.00047}_{-0.00039}$}
\newcommand{\artabfourthree}{$5.482^{+0.019}_{-0.026}$}
\newcommand{\deltatabfourthree}{$0.02355^{+0.00015}_{-0.00012}$}
\newcommand{\deltartabfourthree}{$0.03043^{+0.00086}_{-0.00081}$}
\newcommand{\deltaztabfourthree}{$0.02782^{+0.00050}_{-0.00047}$}
\newcommand{\deltattabfourthree}{$0.02869^{+0.00075}_{-0.00072}$}
\newcommand{\tautabfourthree}{$0.01220^{+0.00018}_{-0.00013}$}
\newcommand{\tonefourtabfourthree}{$0.08992\pm0.00014$}
\newcommand{\tfwhmtabfourthree}{$0.07770^{+0.00017}_{-0.00018}$}
\newcommand{\ruonetabfourthree}{$0.460\pm0.041$}
\newcommand{\rutwotabfourthree}{$0.151^{+0.075}_{-0.076}$}
\newcommand{\zuonetabfourthree}{$0.313\pm0.025$}
\newcommand{\zutwotabfourthree}{$0.218^{+0.056}_{-0.058}$}
\newcommand{\tuonetabfourthree}{$0.365\pm0.040$}
\newcommand{\tutwotabfourthree}{$0.139^{+0.074}_{-0.075}$}
\newcommand{\rptabfourfour}{$1.339^{+0.023}_{-0.021}$}
\newcommand{\mptabfourfour}{$1.207^{+0.073}_{-0.071}$}
\newcommand{\atabfourfour}{$0.02284^{+0.00040}_{-0.00037}$}
\newcommand{\teqtabfourfour}{$1637^{+20}_{-19}$}
\newcommand{\tcirctabfourfour}{$0.00279\pm0.00018$}
\newcommand{\rhoptabfourfour}{$0.623\pm0.034$}
\newcommand{\loggptabfourfour}{$3.222^{+0.021}_{-0.023}$}
\newcommand{\favetabfourfour}{$1.629^{+0.081}_{-0.075}$}
\newcommand{\msinitabfourfour}{$1.206^{+0.073}_{-0.071}$}
\newcommand{\periodtabfourfour}{$1.338231513\pm0.000000019$}
\newcommand{\tctabfourfour}{$2454697.797524\pm0.000023$}
\newcommand{\tttabfourfour}{$2454697.797524\pm0.000023$}
\newcommand{\tstabfourfour}{$2454698.466640\pm0.000023$}
\newcommand{\tzerotabfourfour}{$2455329.442798\pm0.000022$}
\newcommand{\itabfourfour}{$88.89^{+0.65}_{-0.51}$}
\newcommand{\btabfourfour}{$0.106^{+0.048}_{-0.062}$}
\newcommand{\ktabfourfour}{$240\pm12$}
\newcommand{\ptabfourfour}{$0.15348^{+0.00046}_{-0.00039}$}
\newcommand{\artabfourfour}{$5.482^{+0.019}_{-0.026}$}
\newcommand{\deltatabfourfour}{$0.02355^{+0.00014}_{-0.00012}$}
\newcommand{\deltartabfourfour}{$0.03043^{+0.00086}_{-0.00082}$}
\newcommand{\deltaztabfourfour}{$0.02782^{+0.00049}_{-0.00046}$}
\newcommand{\deltattabfourfour}{$0.02868^{+0.00076}_{-0.00071}$}
\newcommand{\tautabfourfour}{$0.01220^{+0.00017}_{-0.00012}$}
\newcommand{\tonefourtabfourfour}{$0.08992\pm0.00014$}
\newcommand{\tfwhmtabfourfour}{$0.07770^{+0.00017}_{-0.00018}$}
\newcommand{\ruonetabfourfour}{$0.460^{+0.041}_{-0.042}$}
\newcommand{\rutwotabfourfour}{$0.152^{+0.076}_{-0.075}$}
\newcommand{\zuonetabfourfour}{$0.313^{+0.025}_{-0.024}$}
\newcommand{\zutwotabfourfour}{$0.219^{+0.055}_{-0.057}$}
\newcommand{\tuonetabfourfour}{$0.365\pm0.040$}
\newcommand{\tutwotabfourfour}{$0.140^{+0.074}_{-0.076}$}
\begin{turnpage}
  \begin{table*}
     \centering
     \caption{Median values and 68\% confidence interval for the WASP-4 host star}
     \label{tab:wasp4b}
     \begin{tabular}{llcccc}
	    \toprule
     \colhead{~~~Parameter} & \colhead{Units} &\colhead{Transit}&\colhead{Transit+SED}&\colhead{MIST+Transit}&\colhead{MIST+Transit+SED}\\
     \toprule
       ~~~~$P$\dotfill&Period (days)\dotfill&\periodtabfourone&\periodtabfourtwo&\periodtabfourthree&\periodtabfourfour\\
       ~~~~$R_P$\dotfill&Radius (\rj)\dotfill&\rptabfourone&\rptabfourtwo&\rptabfourthree&\rptabfourfour\\
       ~~~~$M_P$\dotfill&Mass (\mj)\dotfill&\mptabfourone&\mptabfourtwo&\mptabfourthree&\mptabfourfour\\
       ~~~~$T_C$\dotfill&Time of conjunction$^4$ (\bjdtdb)\dotfill&\tctabfourone&\tctabfourtwo&\tctabfourthree&\tctabfourfour\\
       ~~~~$T_T$\dotfill&Time of min proj sep$^5$ (\bjdtdb)\dotfill&\tttabfourone&\tttabfourtwo&\tttabfourthree&\tttabfourfour\\
       ~~~~$T_0$\dotfill&Optimal conj Time$^6$ (\bjdtdb)\dotfill&\tzerotabfourone&\tzerotabfourtwo&\tzerotabfourthree&\tzerotabfourfour\\
       ~~~~$a$\dotfill&Semi-major axis (AU)\dotfill&\atabfourone&\atabfourtwo&\atabfourthree&\atabfourfour\\
       ~~~~$i$\dotfill&Inclination (Degrees)\dotfill&\itabfourone&\itabfourtwo&\itabfourthree&\itabfourfour\\
       ~~~~$T_{\rm eq}$\dotfill&Equilibrium temperature$^7$ (K)\dotfill&\teqtabfourone&\teqtabfourtwo&\teqtabfourthree&\teqtabfourfour\\
       ~~~~$\tau_{\rm circ}$\dotfill&Tidal circ timescale (Gyr)\dotfill&\tcirctabfourone&\tcirctabfourtwo&\tcirctabfourthree&\tcirctabfourfour\\
       ~~~~$K$\dotfill&RV semi-amplitude (m/s)\dotfill&\ktabfourone&\ktabfourtwo&\ktabfourthree&\ktabfourfour\\
       ~~~~$R_P/R_*$\dotfill&Radius of planet in stellar radii\dotfill&\ptabfourone&\ptabfourtwo&\ptabfourthree&\ptabfourfour\\
       ~~~~$a/R_*$\dotfill&Semi-major axis in stellar radii\dotfill&\artabfourone&\artabfourtwo&\artabfourthree&\artabfourfour\\
       ~~~~$\delta$\dotfill&$(R_P/R_*)^2$\dotfill&\deltatabfourone&\deltatabfourtwo&\deltatabfourthree&\deltatabfourfour\\
       ~~~~$\delta_{\rm R}$\dotfill&Transit depth in R (fraction)\dotfill&\deltartabfourone&\deltartabfourtwo&\deltartabfourthree&\deltartabfourfour\\
       ~~~~$\delta_{\rm z'}$\dotfill&Transit depth in z' (fraction)\dotfill&\deltaztabfourone&\deltaztabfourtwo&\deltaztabfourthree&\deltaztabfourfour\\
       ~~~~$\delta_{\rm TESS}$\dotfill&Transit depth in TESS (fraction)\dotfill&\deltattabfourone&\deltattabfourtwo&\deltattabfourthree&\deltattabfourfour\\
       ~~~~$\tau$\dotfill&Ingress/egress transit duration (days)\dotfill&\tautabfourone&\tautabfourtwo&\tautabfourthree&\tautabfourfour\\
       ~~~~$T_{14}$\dotfill&Total transit duration (days)\dotfill&\tonefourtabfourone&\tonefourtabfourtwo&\tonefourtabfourthree&\tonefourtabfourfour\\
       ~~~~$T_{FWHM}$\dotfill&FWHM transit duration (days)\dotfill&\tfwhmtabfourone&\tfwhmtabfourtwo&\tfwhmtabfourthree&\tfwhmtabfourfour\\
       ~~~~$b$\dotfill&Transit impact parameter\dotfill&\btabfourone&\btabfourtwo&\btabfourthree&\btabfourfour\\
       ~~~~$\rho_P$\dotfill&Density (cgs)\dotfill&\rhoptabfourone&\rhoptabfourtwo&\rhoptabfourthree&\rhoptabfourfour\\
       ~~~~$logg_P$\dotfill&Surface gravity (cgs)\dotfill&\loggptabfourone&\loggptabfourtwo&\loggptabfourthree&\loggptabfourfour\\
       ~~~~$\fave$\dotfill&Incident Flux (\fluxcgs)\dotfill&\favetabfourone&\favetabfourtwo&\favetabfourthree&\favetabfourfour\\
       ~~~~$T_S$\dotfill&Time of eclipse (\bjdtdb)\dotfill&\tstabfourone&\tstabfourtwo&\tstabfourthree&\tstabfourfour\\
       ~~~~$M_P\sin{i}$\dotfill&Minimum mass (\mj)\dotfill&\msinitabfourone&\msinitabfourtwo&\msinitabfourthree&\msinitabfourfour\\
       ~~~~$u_1$\dotfill&R linear limb-darkening coeff\dotfill&\ruonetabfourone&\ruonetabfourtwo&\ruonetabfourthree&\ruonetabfourfour\\
       ~~~~$u_2$\dotfill&R quadratic limb-darkening coeff\dotfill&\rutwotabfourone&\rutwotabfourtwo&\rutwotabfourthree&\rutwotabfourfour\\
       ~~~~$u_1$\dotfill&z' linear limb-darkening coeff\dotfill&\zuonetabfourone&\zuonetabfourtwo&\zuonetabfourthree&\zuonetabfourfour\\
       ~~~~$u_2$\dotfill&z' quadratic limb-darkening coeff\dotfill&\zutwotabfourone&\zutwotabfourtwo&\zutwotabfourthree&\zutwotabfourfour\\
       ~~~~$u_1$\dotfill&TESS linear limb-darkening coeff\dotfill&\tuonetabfourone&\tuonetabfourtwo&\tuonetabfourthree&\tuonetabfourfour\\
       ~~~~$u_2$\dotfill&TESS quadratic limb-darkening coeff\dotfill&\tutwotabfourone&\tutwotabfourtwo&\tutwotabfourthree&\tutwotabfourfour\\
    \hline
    \end{tabular}
    \tablenotetext{}{See Table 3 in \citet{Eastman:2019} for a detailed description of all parameters}
    \tablenotetext{4}{Time of conjunction is commonly reported as the "transit time."}
    \tablenotetext{5}{Time of minimum projected separation is a more correct "transit time."}
    \tablenotetext{6}{Optimal time of conjunction minimizes the covariance between $T_C$ and period}
    \tablenotetext{7}{Assumes no albedo and perfect redistribution}
  \end{table*}
\end{turnpage}
\clearpage

\clearpage
\newcommand{\rptabfiveone}{$1.345^{+0.069}_{-0.070}$}
\newcommand{\mptabfiveone}{$1.19^{+0.14}_{-0.12}$}
\newcommand{\atabfiveone}{$0.0229\pm0.0012$}
\newcommand{\teqtabfiveone}{$1639\pm62$}
\newcommand{\tcirctabfiveone}{$0.00281^{+0.00046}_{-0.00039}$}
\newcommand{\rhoptabfiveone}{$0.609^{+0.12}_{-0.099}$}
\newcommand{\loggptabfiveone}{$3.214^{+0.061}_{-0.062}$}
\newcommand{\favetabfiveone}{$1.64^{+0.26}_{-0.23}$}
\newcommand{\msinitabfiveone}{$1.19^{+0.14}_{-0.12}$}
\newcommand{\rptabfivetwo}{$1.330^{+0.073}_{-0.061}$}
\newcommand{\mptabfivetwo}{$1.199^{+0.078}_{-0.074}$}
\newcommand{\atabfivetwo}{$0.0227^{+0.0012}_{-0.0010}$}
\newcommand{\teqtabfivetwo}{$1643^{+55}_{-57}$}
\newcommand{\tcirctabfivetwo}{$0.00277^{+0.00027}_{-0.00025}$}
\newcommand{\rhoptabfivetwo}{$0.630^{+0.094}_{-0.088}$}
\newcommand{\loggptabfivetwo}{$3.224^{+0.044}_{-0.046}$}
\newcommand{\favetabfivetwo}{$1.66^{+0.23}_{-0.22}$}
\newcommand{\msinitabfivetwo}{$1.198^{+0.078}_{-0.074}$}
\newcommand{\rptabfivethree}{$1.330^{+0.066}_{-0.057}$}
\newcommand{\mptabfivethree}{$1.19^{+0.14}_{-0.12}$}
\newcommand{\atabfivethree}{$0.02268^{+0.0011}_{-0.00097}$}
\newcommand{\teqtabfivethree}{$1642^{+36}_{-38}$}
\newcommand{\tcirctabfivethree}{$0.00278^{+0.00045}_{-0.00039}$}
\newcommand{\rhoptabfivethree}{$0.627^{+0.11}_{-0.097}$}
\newcommand{\loggptabfivethree}{$3.222\pm0.059$}
\newcommand{\favetabfivethree}{$1.65\pm0.15$}
\newcommand{\msinitabfivethree}{$1.19^{+0.14}_{-0.12}$}
\newcommand{\rptabfivefour}{$1.323^{+0.053}_{-0.047}$}
\newcommand{\mptabfivefour}{$1.213^{+0.078}_{-0.074}$}
\newcommand{\atabfivefour}{$0.02257^{+0.00090}_{-0.00081}$}
\newcommand{\teqtabfivefour}{$1646\pm31$}
\newcommand{\tcirctabfivefour}{$0.00271^{+0.00027}_{-0.00025}$}
\newcommand{\rhoptabfivefour}{$0.648^{+0.087}_{-0.082}$}
\newcommand{\loggptabfivefour}{$3.234^{+0.042}_{-0.044}$}
\newcommand{\favetabfivefour}{$1.67^{+0.13}_{-0.12}$}
\newcommand{\msinitabfivefour}{$1.213^{+0.078}_{-0.074}$}
\newcommand{\periodref}{$1.338231512\pm0.000000019$}
\newcommand{\tcref}{$2454697.797524\pm0.000024$}
\newcommand{\ttref}{$2454697.797524\pm0.000024$}
\newcommand{\tsref}{$2454698.466639\pm0.000024$}
\newcommand{\tzeroref}{$2455337.472187\pm0.000022$}
\newcommand{\iref}{$88.87^{+0.67}_{-0.51}$}
\newcommand{\bref}{$0.108^{+0.048}_{-0.064}$}
\newcommand{\kref}{$240\pm12$}
\newcommand{\pref}{$0.15348^{+0.00047}_{-0.00039}$}
\newcommand{\arref}{$5.481^{+0.020}_{-0.026}$}
\newcommand{\deltaref}{$0.02356^{+0.00014}_{-0.00012}$}
\newcommand{\deltarref}{$0.03042^{+0.00086}_{-0.00081}$}
\newcommand{\deltazref}{$0.02781^{+0.00049}_{-0.00046}$}
\newcommand{\deltatref}{$0.02867^{+0.00076}_{-0.00072}$}
\newcommand{\tauref}{$0.01220^{+0.00017}_{-0.00013}$}
\newcommand{\tonefourref}{$0.08992\pm0.00014$}
\newcommand{\tfwhmref}{$0.07770^{+0.00017}_{-0.00018}$}
\newcommand{\ruoneref}{$0.460\pm0.041$}
\newcommand{\rutworef}{$0.152^{+0.075}_{-0.076}$}
\newcommand{\zuoneref}{$0.312^{+0.025}_{-0.024}$}
\newcommand{\zutworef}{$0.219^{+0.055}_{-0.057}$}
\newcommand{\tuoneref}{$0.364^{+0.041}_{-0.040}$}
\newcommand{\tutworef}{$0.141^{+0.075}_{-0.076}$}
\begin{turnpage}
  \begin{table*}
     \centering
     \caption{Median values and 68\% confidence interval for WASP-4b planet fits, separating the transit and stellar models. Values that span all columns are taken from the transit-only fit.}
     \label{tab:wasp4b_notransit}
     \begin{tabular}{llcccc}
	    \toprule
     \colhead{~~~Parameter} & \colhead{Units} &\colhead{None}&\colhead{MIST}&\colhead{SED}&\colhead{MIST+SED}\\
     \toprule
       ~~~~$P$\dotfill&Period (days)\dotfill&\multicolumn{4}{c}{\periodref}\\
       ~~~~$R_P$\dotfill&Radius (\rj)\dotfill&\rptabfiveone&\rptabfivetwo&\rptabfivethree&\rptabfivefour\\
       ~~~~$M_P$\dotfill&Mass (\mj)\dotfill&\mptabfiveone&\mptabfivetwo&\mptabfivethree&\mptabfivefour\\
       ~~~~$T_C$\dotfill&Time of conjunction$^4$ (\bjdtdb)\dotfill&\multicolumn{4}{c}{\tcref}\\
       ~~~~$T_T$\dotfill&Time of min proj sep$^5$ (\bjdtdb)\dotfill&\multicolumn{4}{c}{\ttref}\\
       ~~~~$T_0$\dotfill&Optimal conj Time$^6$ (\bjdtdb)\dotfill&\multicolumn{4}{c}{\tzeroref}\\
       ~~~~$a$\dotfill&Semi-major axis (AU)\dotfill&\atabfiveone&\atabfivetwo&\atabfivethree&\atabfivefour\\
       ~~~~$i$\dotfill&Inclination (Degrees)\dotfill&\multicolumn{4}{c}{\iref}\\
       ~~~~$T_{\rm eq}$\dotfill&Equilibrium temperature$^7$ (K)\dotfill&\teqtabfiveone&\teqtabfivetwo&\teqtabfivethree&\teqtabfivefour\\
       ~~~~$\tau_{\rm circ}$\dotfill&Tidal circ timescale (Gyr)\dotfill&\tcirctabfiveone&\tcirctabfivetwo&\tcirctabfivethree&\tcirctabfivefour\\
       ~~~~$K$\dotfill&RV semi-amplitude (m/s)\dotfill&\multicolumn{4}{c}{\kref}\\
       ~~~~$R_P/R_*$\dotfill&Radius of planet in stellar radii\dotfill&\multicolumn{4}{c}{\pref}\\
       ~~~~$a/R_*$\dotfill&Semi-major axis in stellar radii\dotfill&\multicolumn{4}{c}{\arref}\\
       ~~~~$\delta$\dotfill&$(R_P/R_*)^2$\dotfill&\multicolumn{4}{c}{\deltaref}\\
       ~~~~$\delta_{\rm R}$\dotfill&Transit depth in R (fraction)\dotfill&\multicolumn{4}{c}{\deltarref}\\
       ~~~~$\delta_{\rm z'}$\dotfill&Transit depth in z' (fraction)\dotfill&\multicolumn{4}{c}{\deltazref}\\
       ~~~~$\delta_{\rm TESS}$\dotfill&Transit depth in TESS (fraction)\dotfill&\multicolumn{4}{c}{\deltatref}\\
       ~~~~$\tau$\dotfill&Ingress/egress transit duration (days)\dotfill&\multicolumn{4}{c}{\tauref}\\
       ~~~~$T_{14}$\dotfill&Total transit duration (days)\dotfill&\multicolumn{4}{c}{\tonefourref}\\
       ~~~~$T_{FWHM}$\dotfill&FWHM transit duration (days)\dotfill&\multicolumn{4}{c}{\tfwhmref}\\
       ~~~~$b$\dotfill&Transit impact parameter\dotfill&\multicolumn{4}{c}{\bref}\\
       ~~~~$\rho_P$\dotfill&Density (cgs)\dotfill&\rhoptabfiveone&\rhoptabfivetwo&\rhoptabfivethree&\rhoptabfivefour\\
       ~~~~$logg_P$\dotfill&Surface gravity (cgs)\dotfill&\loggptabfiveone&\loggptabfivetwo&\loggptabfivethree&\loggptabfivefour\\
       ~~~~$\fave$\dotfill&Incident Flux (\fluxcgs)\dotfill&\favetabfiveone&\favetabfivetwo&\favetabfivethree&\favetabfivefour\\
       ~~~~$T_S$\dotfill&Time of eclipse (\bjdtdb)\dotfill&\multicolumn{4}{c}{\tsref}\\
       ~~~~$M_P\sin{i}$\dotfill&Minimum mass (\mj)\dotfill&\msinitabfiveone&\msinitabfivetwo&\msinitabfivethree&\msinitabfivefour\\
       ~~~~$u_1$\dotfill&R linear limb-darkening coeff\dotfill&\multicolumn{4}{c}{\ruoneref}\\
       ~~~~$u_2$\dotfill&R quadratic limb-darkening coeff\dotfill&\multicolumn{4}{c}{\rutworef}\\
       ~~~~$u_1$\dotfill&z' linear limb-darkening coeff\dotfill&\multicolumn{4}{c}{\zuoneref}\\
       ~~~~$u_2$\dotfill&z' quadratic limb-darkening coeff\dotfill&\multicolumn{4}{c}{\zutworef}\\
       ~~~~$u_1$\dotfill&TESS linear limb-darkening coeff\dotfill&\multicolumn{4}{c}{\tuoneref}\\
       ~~~~$u_2$\dotfill&TESS quadratic limb-darkening coeff\dotfill&\multicolumn{4}{c}{\tutworef}\\
    \hline
    \end{tabular}
    \tablenotetext{}{See Table 3 in \citet{Eastman:2019} for a detailed description of all parameters.}
    \tablenotetext{4}{Time of conjunction is commonly reported as the "transit time."}
    \tablenotetext{5}{Time of minimum projected separation is a more correct "transit time."}
    \tablenotetext{6}{Optimal time of conjunction minimizes the covariance between $T_C$ and period.}
    \tablenotetext{7}{Assumes no albedo and perfect redistribution.}
  \end{table*}
\end{turnpage}
\clearpage

\subsection{Tidal Circularization}
\label{sec:tidalcirc}

In general, the eccentricity of hot Jupiters is likely not exactly zero, but ignoring the difference between its actual eccentricity and 0 introduces a negligible error in \rhostar \ compared to our 2\% goal (beyond which the measurement is dominated by \mstar \ systematics). 

One could reasonably argue that we should use the observational constraints on eccentricity such that the presumption of circularity does not bias our measurement of \rhostar. However, for WASP-4 and many hot Jupiters, the observational constraints on the eccentricity are poor and do not account for the strong theoretical expectation we have for tidal circularization. Therefore, the observational limits represent a very conservative upper limit on the allowed eccentricity, which translates to an unnecessarily conservative uncertainty on \rhostar.

\citet{Wang:2011} explored the eccentricity distribution of such short-period planets, but their sample only had a single planet with a period comparable to WASP-4 (HD41004B), and its eccentricity is consistent with zero $\left(e=0.058^{+0.051}_{-0.058}\right)$. They had no planets with a tidal circularization timescale comparable to the 2.8 Myr we compute for WASP-4b, but all the planets in their sample with a tidal circularization timescale of less than 1 Gyr had an eccentricity consistent with zero.

Ultimately, our goal is to compute the most precise and accurate stellar parameters, and to do that, we believe that the theoretical expectation of tidal circularization (at least in the case of WASP-4b) is more reliable than our theoretical understanding for stellar evolution. However, we should be clear that it is still generally useful to fit for eccentricity so that we can test the theoretical expectations of tidal circularization, rather than assume it as we do here. 

\section{Discussion}
\label{sec:discussion}

Among the set of 1145 default (\texttt{DEFAULT\_FLAG=1}) transiting planets (\texttt{TRAN\_FLAG=1}) in the exoplanet archive where the stellar density and its uncertainty are populated, 56 have host stars with a fractional \rhostar \ uncertainty less than 2\%. It grows to 426 systems if we use a threshold of 9\% (where we can beat the systematic floor in the SED/MIST-derived \logg), 503 systems if we use a threshold of 10.3\% (where we can beat the systematic floor in the SED/MIST-derived \rstar), and 556 systems if we use a threshold of 11.5\% (where we can beat the systematic floor in the spectroscopic \teff). 

These are likely a significant undercount of the number of systems suitable to such precision given the heterogeneity of the sample, the relative rarity of simultaneous modeling of the star and planet, and the fact that only 30\% of the default set of transiting planets even have stellar densities populated. It is possible that some of these densities are optimistically derived from evolutionary models while ignoring systematic errors rather than a transit light curve. However, only 6\% of nontransiting planets have quoted stellar densities compared to 30\% of transiting planets, implying that the transit was used for most when available.

Regardless, this technique could likely be applied to a significant fraction of transiting planet hosts to improve the  stellar and planetary parameters. Even nontransiting planet hosts are likely to see improved \logg \ precision, as described in \S \ref{sec:loggfrommr}, which can be used to improve spectroscopic measurements of \teff \ and \feh. While a precision similar to what we achieve here is commonly reported in the literature, few have accounted for the systematic uncertainties in the stellar parameters shown by \citet{Tayar:2022}, and so they may be too optimistic.

The results shown here emphasize just how important it is to model the star along with the planet to improve the precision of both. It is possible that a large sample of well-measured transit light curves may even help inform stellar models. This method is competitive with gold standard measurements like asterosiesmology or eclipsing binary stars but broadens the pool of applicable stars dramatically. This, in turn, could give us a precise probe into stellar parameters that enable us to test and refine the evolutionary models. Because our derived parameters are still limited by systematics in the stellar models, further improvement in stellar models would yield additional refinement with currently known stellar densities. It may even be fruitful to explore how best to take advantage of the precise stellar density and bolometric flux constraints when constructing the evolutionary models themselves, as these are directly and precisely measured for a much larger sample of stars than are typically used to anchor stellar models.

\acknowledgements
Work by J.D.E. was funded by NASA ADAP 80NSSC19K1014. Computations in this paper were run on the Cannon cluster supported by the FAS Division of Science, Research Computing Group at Harvard University.

This research has made use of the NASA Exoplanet Archive, which is operated by the California Institute of Technology, under contract with the National Aeronautics and Space Administration under the Exoplanet Exploration Program.

This work has made use of data from the European Space Agency (ESA) mission
{\it Gaia} (\url{https://www.cosmos.esa.int/gaia}), processed by the {\it Gaia}
Data Processing and Analysis Consortium (DPAC,
\url{https://www.cosmos.esa.int/web/gaia/dpac/consortium}). Funding for the DPAC
has been provided by national institutions, in particular the institutions
participating in the {\it Gaia} Multilateral Agreement.

This publication makes use of data products from the Two Micron All Sky Survey, which is a joint project of the University of Massachusetts and the Infrared Processing and Analysis Center/California Institute of Technology, funded by the National Aeronautics and Space Administration and the National Science Foundation.

This publication makes use of data products from the Wide-field Infrared Survey Explorer, which is a joint project of the University of California, Los Angeles, and the Jet Propulsion Laboratory/California Institute of Technology, funded by the National Aeronautics and Space Administration.

We thank the referee for thoughtful comments that have improved the clarity of the manuscript.

\facilities {Exoplanet Archive, Gaia, FLWO:2MASS, CTIO:2MASS, WISE}




\end{document}